\documentclass[a4paper,11pt]{article}

\usepackage{jheppub}
\usepackage[T1]{fontenc}
\usepackage[utf8]{inputenc}
\usepackage{slashed}
\usepackage{subcaption}
\usepackage{todonotes}
\usepackage{hyperref}
\usepackage{bm}

\usepackage{tikz}
\usetikzlibrary{snakes}
\usetikzlibrary{decorations}
\usetikzlibrary{trees}
\usetikzlibrary{decorations.pathmorphing}
\usetikzlibrary{decorations.markings}
\usetikzlibrary{external}
\usetikzlibrary{intersections}
\usetikzlibrary{shapes,arrows}
\usetikzlibrary{arrows.meta}
\usetikzlibrary{calc}
\usetikzlibrary{shapes.misc}
\usetikzlibrary{decorations.text}
\usetikzlibrary{backgrounds}
\usetikzlibrary{fadings}
\usetikzlibrary{tikzmark,calc,arrows,shapes,decorations.pathreplacing}

\tikzset{
	graviton/.style={decorate,line width=0.15mm, 
	decoration={snake,amplitude=.6mm, segment length=1.5mm}
	},
	photon/.style={decorate, decoration={snake}, draw=red},
	scalar/.style={postaction={decorate},
	},
	massive/.style={postaction={decorate},
		line width=0.75mm,
	},
	massless/.style={postaction={decorate},
	},
	masslessWithDot/.style={postaction={decorate},
		decoration={
			markings,
			mark=at position 0.5 with {\fill circle (2pt);}}
	},
	massiveWithDot/.style={postaction={decorate},
		line width=0.5mm,
		decoration={
			markings,
			mark=at position 0.5 with {\fill circle (2pt);}}
	},
	massiveWithArrow/.style={postaction={decorate},
		line width=0.75mm,
		decoration={
			markings,
			mark=at position 0.5 with {\arrow{latex}}}
	},
	massiveWithArrowB/.style={postaction={decorate},
		line width=0.75mm,
		decoration={
			markings,
			mark=at position 0.95 with {\arrow{latex}}}
	},
	massiveLin/.style={postaction={decorate},
		double,
		thick,
		fill=white
	},
	massivePhi/.style={postaction={decorate},
		line width=0.75mm,
		dashed
	},
	masslessPhi/.style={postaction={decorate},
		dashed
	},
	unitaryCut/.style={postaction={draw,densely dashed,blue,thin},
		line width = 0.2cm,white
	},
	gluon/.style={decorate, draw=magenta,
		decoration={coil,amplitude=4pt, segment length=5pt}},
	partial ellipse/.style args={#1:#2:#3}{
		insert path={+ (#1:#3) arc (#1:#2:#3)}
	},
	cross/.style={cross out, draw=black, minimum size=2*(#1-\pgflinewidth), inner sep=0pt, outer sep=0pt},
	branchCut/.style={postaction={decorate},
		snake=zigzag,
		decoration = {snake=zigzag,segment length = 2mm, amplitude = 2mm}	
	}
	cross/.default={1pt}
}

\colorlet{mred}{black!30!red}
\colorlet{mgreen}{black!30!green}
\colorlet{mblue}{black!30!blue}
\colorlet{morange}{blue!70!red}

\newcommand{\bubCut}{
\begin{tikzpicture}
\coordinate (e2) at (-1.00, 0.75);
		\coordinate (e3) at ( 1.00, 0.75);
		\coordinate (e4) at ( 1.00,-0.75);
		\coordinate (e1) at (-1.00,-0.75);	
		
		\coordinate (v1) at (0,-0.5);
		\coordinate (v2) at (0, 0.5);
		
		\draw[graviton] (e1) -- (v1);
		\draw[graviton] (v1) -- (e4);
		\draw[graviton] (e3) -- (v2);
		\draw[graviton] (e2) -- (v2);
		\draw[massive] (v1) to[out=20,in=-20] (v2);
		\draw[massive] (v2) to[out=160,in=-160] (v1);

        \draw[fill=lightgray, opacity=1] (v1) ellipse (0.4 and 0.2);
        \draw[fill=lightgray, opacity=1] (v2) ellipse (0.4 and 0.2);
\end{tikzpicture}}

\newcommand{\threePtAmpSGG}{
\begin{tikzpicture}
        \coordinate (e1) at (0.5, 0.866025);  
		\coordinate (e2) at (-1.00, 0);
		\coordinate (e3) at (0.5, -0.866025);
		
		\coordinate (v1) at (0,0);  
		
		\draw[photon] (e1) -- (v1);
		\draw[photon] (e2) -- (v1);
		\draw[masslessPhi] (e3) -- (v1);
		
		\node[above right] at (e1) {$G$};
        \node[left]  at (e2) {$G$};
        \node[below right] at (e3) {$S$};
		
        \draw[fill=lightgray, opacity=1] (v1) ellipse (0.2 and 0.2);
\end{tikzpicture}}

\newcommand{\threePtAmpggPhi}{
\begin{tikzpicture}
        \coordinate (e1) at (0.5, 0.866025);  
		\coordinate (e2) at (-1.00, 0);
		\coordinate (e3) at (0.5, -0.866025);
		
		\coordinate (v1) at (0,0);  
		
		\draw[graviton] (e1) -- (v1);
		\draw[graviton] (e2) -- (v1);
		\draw[masslessPhi] (e3) -- (v1);
		
		\node[above right] at (e1) {$h^-$};
        \node[left]  at (e2) {$h^-$};
        \node[below right] at (e3) {$\varphi$};
		
        \draw[fill=lightgray, opacity=1] (v1) ellipse (0.2 and 0.2);
\end{tikzpicture}}

\newcommand{\exchangeSuperSuper}{
\begin{tikzpicture}
        \coordinate (e1) at (-1, -1);  
		\coordinate (e2) at (-1, 1);
		\coordinate (e3) at (1, 1);
		\coordinate (e4) at (1, -1);
		
		\coordinate (v1) at (0,-.75);  
		\coordinate (v2) at (0,.75); 
		
		\draw[photon] (e1) -- (v1);
		\draw[photon] (e4) -- (v1);
		\draw[photon] (e2) -- (v2);
		\draw[photon] (e3) -- (v2);
		\draw[masslessPhi] (v1) -- (v2);
		
		\node[below left] at (e1) {$\overline{G}$};
        \node[above left]  at (e2) {$G$};
        \node[above right] at (e3) {$G$};
        \node[below right] at (e4) {$\overline{G}$};
		
        \draw[fill=lightgray, opacity=1] (v1) ellipse (0.2 and 0.2);
        \draw[fill=lightgray, opacity=1] (v2) ellipse (0.2 and 0.2);
\end{tikzpicture}}

\newcommand{\exchangeKahlerKahler}{
\begin{tikzpicture}
        \coordinate (e1) at (-1, -1);  
		\coordinate (e2) at (-1, 1);
		\coordinate (e3) at (1, 1);
		\coordinate (e4) at (1, -1);
		
		\coordinate (v1) at (0,-.75);  
		\coordinate (v2) at (0,.75); 
		
		\draw[photon] (e1) -- (v1);
		\draw[photon] (e4) -- (v1);
		\draw[photon] (e2) -- (v2);
		\draw[photon] (e3) -- (v2);
		\draw[masslessPhi] (v1) -- (v2);
		
		\node[below left] at (e1) {$\overline{G}$};
        \node[above left]  at (e2) {$G$};
        \node[above right] at (e3) {$G$};
        \node[below right] at (e4) {$\overline{G}$};
        
        \node[rectangle, fill=lightgray, opacity=1, draw, minimum width = .4cm, minimum height = .4cm] at (v1) {};
        \node[rectangle, fill=lightgray, opacity=1, draw, minimum width = .4cm, minimum height = .4cm] at (v2) {};
\end{tikzpicture}}

\newcommand{\contactKahler}{
\begin{tikzpicture}
        \coordinate (e1) at (-1, -1);  
		\coordinate (e2) at (-1, 1);
		\coordinate (e3) at (1, 1);
		\coordinate (e4) at (1, -1);
		
		\coordinate (v1) at (0,0);  
		
		\draw[photon] (e1) -- (v1);
		\draw[photon] (e4) -- (v1);
		\draw[photon] (e2) -- (v1);
		\draw[photon] (e3) -- (v1);
		
		\node[below left] at (e1) {$\overline{G}$};
        \node[above left]  at (e2) {$G$};
        \node[above right] at (e3) {$G$};
        \node[below right] at (e4) {$\overline{G}$};
        
        \node[rectangle,fill=lightgray, opacity=1,draw,minimum width = .4cm, minimum height = .4cm] at (v1) {};
\end{tikzpicture}}

\newcommand{\hmhmhphp}{
\begin{tikzpicture}
        \coordinate (e1) at (-1, -1);  
		\coordinate (e2) at (-1, 1);
		\coordinate (e3) at (1, 1);
		\coordinate (e4) at (1, -1);
		
		\coordinate (v1) at (-.75,0);  
		\coordinate (v2) at (.75,0); 
		
		\draw[graviton] (e1) -- (v1);
		\draw[graviton] (e2) -- (v1);
		\draw[graviton] (e3) -- (v2);
		\draw[graviton] (e4) -- (v2);
		\draw[masslessPhi] (v1) -- (v2);
		
		\node[below left] at (e1) {$1^-$};
        \node[above left]  at (e2) {$2^-$};
        \node[above right] at (e3) {$3^+$};
        \node[below right] at (e4) {$4^+$};
		
        \draw[fill=lightgray, opacity=1] (v1) ellipse (0.2 and 0.2);
        \draw[fill=lightgray, opacity=1] (v2) ellipse (0.2 and 0.2);
\end{tikzpicture}}

\newcommand{\boxInt}{
\begin{tikzpicture}
        \coordinate (e1) at (-1.00,-1.00);  
		\coordinate (e2) at (-1.00, 1.00);
		\coordinate (e3) at ( 1.00, 1.00);
		\coordinate (e4) at ( 1.00,-1.00);
		
		\coordinate (v1) at (-.6,-0.6);  
		\coordinate (v2) at (-.6, 0.6);
		\coordinate (v3) at ( .6, 0.6);
		\coordinate (v4) at ( .6,-0.6);	
		
		\draw[massless] (e1) -- (v1);
		\draw[massless] (e2) -- (v2);
		\draw[massless] (e3) -- (v3);
		\draw[massless] (e4) -- (v4);

		\draw[massive] (v1) -- (v2) -- (v3) -- (v4) -- (v1);

\end{tikzpicture}}

\newcommand{\triInt}{
\begin{tikzpicture}
        \coordinate (e1) at (-1.00,-1.00);  
		\coordinate (e2) at (-1.00, 1.00);
		\coordinate (e3) at ( 1.00, 1.00);
		\coordinate (e4) at ( 1.00,-1.00);
		
		\coordinate (v2) at (-.6, 0.6);
		\coordinate (v3) at ( .6, 0.6);
		\coordinate (v4) at (  0,-0);
		\coordinate (v1) at (  0,-0.6);
		
		\draw[massless] (e1) -- (v1);
		\draw[massless] (e2) -- (v2);
		\draw[massless] (e3) -- (v3);
		\draw[massless] (e4) -- (v1);
		\draw[massless] (v4) -- (v1);

		\draw[massive] (v2) -- (v3) -- (v4) -- (v2);

\end{tikzpicture}}

\newcommand{\bubInt}{
\begin{tikzpicture}
\coordinate (e2) at (-1.00, 0.75);
		\coordinate (e3) at ( 1.00, 0.75);
		\coordinate (e4) at ( 1.00,-0.75);
		\coordinate (e1) at (-1.00,-0.75);	
		
		\coordinate (v3) at (0,-0.6);
		\coordinate (v4) at (0, 0.6);
		\coordinate (v1) at (0,-0.3);
		\coordinate (v2) at (0, 0.3);
		
		\draw[massless] (v1) -- (v3);
		\draw[massless] (v2) -- (v4);
		
		\draw[massless] (e1) -- (v3);
		\draw[massless] (v3) -- (e4);
		\draw[massless] (e3) -- (v4);
		\draw[massless] (e2) -- (v4);
		
	    \draw[line width=0.75mm] (0,0) circle (.3);
\end{tikzpicture}}

\graphicspath{{figures/}}
\usepackage[export]{adjustbox}

\def\id{\protect{{1 \kern-.28em {\rm l}}}}

\newcommand{\nocontentsline}[3]{}
\newcommand{\tocless}[2]{\bgroup\let\addcontentsline=\nocontentsline#1{#2}\egroup}

\newcommand{\ab}[1]{\langle #1 \rangle}

\newcommand{\nn}{\nonumber}
\renewcommand{\imath}{\mathrm{i}}

\begin{document}
\title{Effective Field Theory Islands from Perturbative and Nonperturbative Four-Graviton Amplitudes}

\author[1]{Zvi Bern,}
\affiliation[1]{Mani L. Bhaumik Institute for Theoretical Physics,\\
UCLA Department of Physics and Astronomy, Los Angeles, CA 90095, USA}
\emailAdd{bern@physics.ucla.edu}

\author[1]{Enrico Herrmann,}
\emailAdd{eh10@g.ucla.edu}

\author[1]{Dimitrios Kosmopoulos,}
\emailAdd{dkosmopoulos@physics.ucla.edu}

\author[2]{Radu Roiban}
\affiliation[2]{Institute for Gravitation and the Cosmos,\\
Pennsylvania State University,
University Park, PA 16802, USA
}
\emailAdd{radu@phys.psu.edu}

\abstract{
Theoretical data obtained from physically sensible field and string theory models suggest that gravitational Effective Field Theories (EFTs) live on islands that are tiny compared to current general bounds determined from unitarity, causality, crossing symmetry, and a good high-energy behavior.  In this work, we present explicit perturbative and nonperturbative $2 \to 2$ graviton scattering amplitudes and their associated low-energy expansion in spacetime dimensions $D\geq 4$ to support this notion. Our new results include a first nonperturbative example consisting of a $D=4$, $\mathcal{N}=1$ supersymmetric field theory that is coupled weakly to gravity. We show that this nonperturbative model lies on the same islands identified using four-dimensional perturbative models based on string theory and minimally-coupled matter circulating a loop. Furthermore, we generalize the previous four-dimensional perturbative models based on string theory and minimally-coupled massive spin-0 and spin-1 states circulating in the loop to $D$ dimensions.  Remarkably, we again find that the low-energy EFT coefficients lie on small islands. These results offer a useful guide towards constraining possible extensions of Einstein gravity. 
}


\maketitle

%
\section{Introduction}
\label{sec:intro}
%

The language of effective field theory (EFT) is a widely accepted framework in which to formulate the physical laws at a certain energy scale (often referred to as the infrared (IR) scale). Typically, this language is used in the context of current and near-future high-energy physics experiments~\cite{Buchmuller:2001dc,Brivio:2017vri}, but has also found applications in a variety of topics including hydrodynamics~\cite{Dubovsky:2011sj}, inflation~\cite{Cheung:2007st}, the large scale structure of the Universe~\cite{Carrasco:2012cv}, and the description of binary motion in general relativity \cite{Goldberger:2004jt}, among many others. 
In EFTs, the relevant physics is parameterized by independent local operators that capture all relevant physical degrees of freedom and are consistent with the known symmetries of the problem. Examples of such symmetries include Lorentz invariance and possibly gauge or global symmetries. The unknown physics at high energy or ultraviolet (UV) physics is then systematically parameterized by successively including higher-dimension operators that capture corrections to low-energy observables. Naively, the (Wilson) coefficients of such higher-dimension operators are undetermined and can take on arbitrary values. 
However, desirable properties of the underlying theory such as causality (analyticity) and unitarity impose nontrivial constraints or bounds on the allowed values of the low-energy couplings~\cite{Adams:2006sv,Camanho:2014apa}. A way to expose these bounds is to study the connection of the $2\to 2$ scattering amplitude in the IR and the UV by means of \emph{dispersion relations} which relate low-energy Wilson coefficients to the discontinuities of the UV amplitude by a contour deformation subject to certain assumptions about the Regge growth of amplitudes at large energies in the complex plane, see e.g.~Refs.~\cite{Froissart:1961ux,Martin:1962rt,Maldacena:2015waa,Haring:2022cyf}. In recent years this basic philosophy has been systematized to extract various nontrivial constraints~\cite{Bellazzini:2020cot, Arkani-Hamed:2020blm, Bern:2021ppb, Caron-Huot:2021enk, Caron-Huot:2022ugt,Chiang:2022jep,Chiang:2022ltp}.

It is critical to understand the full implications of these constraints and whether sensible physical theories must necessarily lie in small regions of the EFT parameter space. Ref.~\cite{Bern:2021ppb} observed that the Wilson coefficients of two distinct classes of gravitational effective field theories derived from models of UV physics populate small \emph{theory islands} in the larger space allowed by current dispersive arguments. The first class are string theories, which are ultraviolet complete, and the second class comes from integrating out minimally coupled matter circulating in loops.  In both instances gravity is assumed to be weakly coupled so that only the leading order contributions are required, corresponding to tree level in string theory and one loop in the field-theory models.  While the field-theory models are not full UV completions, they can be interpreted as intermediate-scale theories, which satisfy all the assumptions used to derive bounds on the EFT coefficients.  The fact that such dissimilar models land on the same small theory island suggests that sensible theories should obey much stronger constraints than have been found as yet from the general arguments.
The observed small islands were interpreted as being related to \emph{low spin dominance}---essentially the property that the spectral density in these models is dominated by the lowest spin partial waves.  

While suggestive, an obvious question is whether the appearance of small islands is an artifact of the special theories that were considered or whether they are generic for physically sensible theories. Here we provide evidence towards the latter by obtaining data from two new classes of theories.  The first is a nonperturbative strongly coupled $\mathcal{N}=1$ supersymmetric gauge theory which is then weakly coupled to gravity and the second is matter minimally coupled to gravity in $D>4$ spacetime dimensions. Specifically, we present explicit results in $D=6,10$ dimensions with further data and evaluation routines available in the ancillary files.  The well known string-theory amplitudes in $D>4$ provide a third class of EFT data. We use this data to support the notion that small theory islands are not a special feature of $D=4$ perturbative examples, but indeed generalize beyond the cases analyzed in Ref.~\cite{Bern:2021ppb}. It remains a challenge to find the tightest bounds that physically sensible EFTs must satisfy. Some recent progress on improving bounds is found in Refs.~\cite{Caron-Huot:2022ugt,Chiang:2022jep}.  

One of the key lessons of the modern scattering amplitudes program is to focus on gauge- and field-redefinition-invariant quantities. In this spirit, as in Refs.~\cite{Bellazzini:2020cot, Arkani-Hamed:2020blm, Bern:2021ppb, Caron-Huot:2021enk, Caron-Huot:2022ugt}, we focus on the low-energy expansions of scattering amplitudes directly, rather than Wilson coefficients in a Lagrangian that are subject to field-redefinition and integration-by-parts ambiguities. Assuming that gravity couples weakly, we can work to tree-level accuracy in the EFT. There is then a one-to-one map between S-matrix elements in the IR and Wilson coefficients in any given basis of operators, see e.g.~\cite{Shadmi:2018xan}. In this way, the low-energy amplitude can be schematically expanded in the form\footnote{Below, we denote amplitude coefficients by their monomial term, e.g.~$a[s^{k-q}t^q]$. For the sake of compactness, here we simply use $a_{k,q}$ which will have a different meaning for a particular $4D$ helicity amplitude.} 
\vspace{-.3cm}
\begin{align}
    \mathcal{M}_{{\rm IR}}(s,t) \sim \text{light exchange} 
    + \sum_{k\geq q\geq 0} 
    a_{k,q}\  s^{k-q} t^q\,,
    \label{amplitude1}
\end{align}
where the Mandelstam invariants are $s= (p_1+p_2)^2$, $t=(p_1+p_4)^2$, and $u=(p_1+p_3)^2$. As usual, for massless external states, they satisfy the relation $u=-s-t$. The terms denoted by ``light exchange'' correspond to low-energy \emph{poles} from massless or light (relative to the scale of Mandelstam invariants) exchange of states that are within the low-energy EFT. Finally, the $a_{k,q}$ parameterize new four-point contact interactions graded by mass dimension $k$.

In the ultraviolet, it is convenient to parameterize the unknown physics in terms of the partial-wave expansion of the amplitude, involving the spectral density and some characteristic polynomial of the scattering angle $\cos \theta = 1 + 2t/s$ (for massless external states in the $s$-channel center of mass) that encodes the Poincar\'e-invariance properties akin to conformal partial waves in conformal-field-theory (CFT) correlation functions \cite{Ferrara:1972kab,Ferrara:1974ny,Dolan:2000ut,Dolan:2003hv}. For external scalars, these are the Gegenbauer polynomials (see e.g.~\cite{Arkani-Hamed:2020blm}) in general $D$ and the Wigner-$d$ matrices for spinning external states in $D=4$, see e.g.~Refs.~\cite{Itzykson:1980rh,Hebbar:2020ukp}.

The simplest incarnation of the bounds on the coefficients $a_{k,q}$ in Eq.~\eqref{amplitude1} is relatively easy to understand. In the presence of some 
elastic channel where the `out' state is the same as the `in' state, in the forward limit (i.e. $t\rightarrow 0$) the discontinuity of the amplitude becomes an absolute square which then implies positivity constraints on EFT amplitude coefficients \cite{Adams:2006sv}:
\begin{align}
a_{k,0} \sim \langle {\rm in} | T^\dagger T |{\rm in}\rangle = \big| T| {\rm in}\rangle \big|^2 \geq 0 \ .
\end{align}
Such bounds have first appeared in the context of chiral Lagrangians and pion scattering \cite{Pham:1985cr,Ananthanarayan:1994hf,Pennington:1994kc}, before experiencing a revival inspired by the seminal works of Refs.~\cite{Adams:2006sv,Camanho:2014apa}. Recently, similar bounds \cite{Nicolis:2009qm,Bellazzini:2015cra,deRham:2017avq,deRham:2017zjm,Bellazzini:2020cot,Sinha:2020win,Chowdhury:2021ynh,Bellazzini:2021oaj} were organized into a novel geometric structure termed the EFT-hedron \cite{Arkani-Hamed:2020blm} (see also Ref.~\cite{Chiang:2021ziz,Chiang:2022jep,Bern:2021ppb}), related to the Weak Gravity Conjecture \cite{Cheung:2014ega,Cheung:2014vva,Arkani-Hamed:2021ajd}, the analytic bootstrap in AdS/CFT \cite{Caron-Huot:2020adz,Caron-Huot:2021enk}, and applied to the Standard Model EFT and pion scattering \cite{Distler:2006if,Manohar:2008tc,Remmen:2019cyz,Remmen:2020uze}. Furthermore, these bounds were refined away from the forward limit \cite{Caron-Huot:2020cmc,Caron-Huot:2021rmr,Caron-Huot:2022ugt} in order to handle cases with gravitational couplings where the $t$-channel graviton exchange causes difficulties with some of the naive forward limit bounds. Cases with different external helicity configurations, which individually cannot be considered as elastic scattering, were also considered in Refs.~\cite{Bern:2021ppb,Caron-Huot:2022ugt,Chiang:2022jep}, in a spirit similar to Refs.~\cite{Cheung:2016yqr,Zhang:2020jyn,Li:2021lpe}. A key feature, common to all presently known bounds is the appearance of the demarcation of allowed and disallowed regions in the space of low-energy couplings $a_{k,q}$.

As already noted above, in previous four-dimensional studies, explicit string- and field-theory data suggest that physical EFTs live on small theory islands~\cite{Bern:2021ppb}.  In contrast to the four-dimensional case, and for the $D$-dimensional scattering of scalar particles, the spinning partial-wave decomposition that enter the UV part of the dispersion relations are not presently analyzed for the scattering of $D$-dimensional spinning states (see however our ``note added'' below and the upcoming work of Ref.~\cite{Caron-Huot:2022toAppear}). Nevertheless, independently of the availability of precise bounded regions, we can ask where do explicit data lie in order to guide further explorations. Here, we address the question of whether similar islands are observed for more general models than the ones considered in~\cite{Bern:2021ppb}. We do so by obtaining new explicit examples of UV models, including nonperturbative matter and cases outside of four dimensions, from which we extract the low-energy expansion coefficients for gravitational scattering amplitudes. Our analysis further supports the notion that small theory islands are a robust feature of gravitational EFTs. Our example of nonperturbative matter in gravitational $2\to2$ scattering opens up a new class of possible theories to analyze in the future. Higher dimensions are interesting for various reasons, including that they allow for analyses of bounds that avoid complications with IR singularities~\cite{Caron-Huot:2021rmr} and because 10 dimensions is natural for addressing the question of where does string theory lie in the space of possible UV completions~\cite{Guerrieri:2021ivu}.

The remainder of this paper is organized as follows: In section~\ref{sec:kinematics} we summarize our kinematic conventions for $D$-dimensional graviton scattering including a parametrization of the center-of-mass momenta and the polarization states of the external gravitons. In section~\ref{sec:np_data}, we use supersymmetric arguments to generate a first four-dimensional example of nonperturbative matter for gravitational $2\to2$ scattering. In section~\ref{sec:string_data} we discuss the straightforward case of tree-level four-graviton string-theory amplitudes in general spacetime dimensions. Our results for one-loop minimally-coupled graviton amplitudes in $D$ dimensions,
from which we extract nontrivial EFT data, is presented in section~\ref{sec:FT_data}. We consider the cases of minimally-coupled massive spin-0 and spin-1 matter circulating in the loop. We also briefly summarize the well-known amplitudes techniques such as generalized unitarity and integration tools used to evaluate and manipulate such expressions. In section~\ref{sec:plots} we give a summary of our data by plotting some of the obtained low-energy amplitude coefficients that should serve as a useful guide for any near-future attempts to place dispersive bounds on higher-dimensional graviton scattering. We supply the relevant amplitudes and their low-energy expansion in a computer-readable form as ancillary files to this paper. We close with conclusions and a future outlook in section~\ref{sec:conclusions}.

\section{External kinematics} 
\label{sec:kinematics}

To describe $2\to 2$ graviton scattering in $D$ dimensions  we introduce external momenta $p_i$, where $i=1,\ldots,4$ labels the external graviton in question, and work in an all-incoming convention. To capture scattering of all possible external states, we use formal polarization tensors for the gravitons 
\begin{align}
    \label{Eq:PolTensorFactorized}
    \varepsilon^{\mu\nu}_i = \varepsilon^\mu_i \varepsilon^\nu_i\,,
    \hspace{.2cm} \text{with } \hspace{.2cm}
    p_i\cdot \varepsilon_i =0\,
    \hspace{.2cm} \text{and } \hspace{.2cm}
    \varepsilon_i\cdot \varepsilon_i =0\,.
\end{align}
We express gravitational polarization tensors in terms of transverse, null polarization vectors $\varepsilon^\mu_i$. Note that there are $D-2$ such independent null vectors in $D$ spacetime dimensions, while there are $D(D-3)/2$ independent symmetric traceless tensors. The expressions in terms of the above tensors capture the entire space of states for the $D$-dimensional gravitons. Indeed, a generic polarization tensor $E^{\mu \nu} \equiv \varepsilon^\mu \tilde{\varepsilon}^\nu + \tilde{\varepsilon}^\mu \varepsilon^\nu$ may always be written in terms of linear combinations of factorized tensors, e.g.~\cite{Boels:2009bv,Chowdhury:2019kaq}
\begin{align}
 \label{Eq:PolTensorGeneric}
 \varepsilon^\mu \tilde{\varepsilon}^\nu + \tilde{\varepsilon}^\mu \varepsilon^\nu
 =(\varepsilon+\tilde{\varepsilon})^\mu (\varepsilon+\tilde{\varepsilon})^\nu  -\varepsilon^\mu  \varepsilon^\nu 
 -\tilde{\varepsilon}^\mu \tilde{\varepsilon}^\nu
 \,,
\end{align}
where we take $\varepsilon \cdot \tilde{\varepsilon}=0$ to ensure tracelessness.

To study specific examples, we introduce explicit momenta and polarization tensors. We consider scattering in the center-of-mass frame, where
\begin{alignat}{3}
p^\mu_1 & = \sqrt{\frac{s}{2}}
\begin{pmatrix}+1\\-1 \\0\\\vec{0}_{D{-}3}\end{pmatrix} ,
\hskip 1.5 cm 
&& p^\mu_2 &&= \sqrt{\frac{s}{2}}
\begin{pmatrix}+1\\+1\\0\\ \vec{0}_{D{-}3}\end{pmatrix} ,
\nn \\[9pt]
\hskip 1.5 cm
p^\mu_3 &= \sqrt{\frac{s}{2}}
\begin{pmatrix}-1\\-\cos \theta\\-\sin \theta\\\vec{0}_{D{-}3}\end{pmatrix} ,
&& p^\mu_4 &&= \sqrt{\frac{s}{2}}
\begin{pmatrix}-1\\\cos \theta\\\sin \theta\\\vec{0}_{D{-}3}\end{pmatrix} ,
\end{alignat}
and the scattering angle $\theta$ is related to the Mandelstam invariants via $\cos\theta = 1+\frac{2t}{s}$, with $s>0$, and $-s<t<0$ for physical $s$-channel scattering. 
In all examples analyzed in this paper, we consider external polarization tensors of the factorized form in Eq.~(\ref{Eq:PolTensorFactorized}). Different cases can also be obtained straightforwardly as explained above. Focusing on even $D$, given a set of spatial unit vectors $e_{a}^\mu = \delta_{a}^\mu$, with $a=1,\ldots,D$, we define (see e.g.~Ref.~\cite{Boels:2009bv})
\begin{align}
\label{eq:pol_def}
    \varepsilon_{1,2n^\pm}^\mu = \frac{1}{\sqrt{2}} (e^\mu_{2n-1} \pm \imath\,  e^\mu_{2n})\,, \hskip 1 cm  n = 2, \ldots \frac{D}{2}\,.
\end{align}
We obtain the polarization vectors for the other three gravitons using appropriate rotations. More details on the polarization choices are included in the ancillary files with explicit evaluation code for all Lorentz products for graviton polarizations similar to the ones discussed here. Note that for $D=4$ ($n=2$) the polarizations in Eq.~(\ref{eq:pol_def}) describe helicity states. Indeed, for this choice our results reproduce the ones obtained in Ref.~\cite{Bern:2021ppb} using spinor-helicity methods.

\section{Non-perturbative data} 
\label{sec:np_data}

To gain some direct indication on nonperturbative low-energy effective actions, we consider a matter-coupled ${\cal N}=1$ supersymmetric gauge theory which we couple to gravity. 
This gauge theory confines in flat space at some scale $\Lambda$, whose specific relation to the high-energy couplings will not be important. We assume that $\Lambda$ is relatively high and that the the low-energy theory is described by the glueball superfield which is much lighter than the confinement scale. This can be arranged by adjusting the couplings of the high-energy theory. 
We then focus our discussion on energies below the mass $m_s$ of the glueball superfield.
Thus we can ignore terms ${\cal O}(p^2/\Lambda^2)$, but we cannot ignore terms ${\cal O}(p^n/m_S^n)$.

The Wilsonian effective action below the confinement scale for an ${\cal N}=1$ supersymmetric gauge theory with (holomorphic) tree-level superpotential $W_\text{tree}(\phi, Q, {\bar Q})$ depending on some chiral superfields $\phi$ in the adjoint representation and other superfields $Q, {\bar Q}$ in the fundamental representation, has the standard form
\begin{equation}
\hspace{-.4cm}
{\cal L} = \int d^4\theta\, K_\text{eff}(S, {\bar S}, G, {\bar G}) 
         + \int d^2\theta\, W_\text{eff}(S, G) 
         + \text{h.c.} \, ,
\hspace{-.3cm}         
\label{Lagrangian}
\end{equation}
where $S$ is the glueball (chiral) superfield, $G$ is the Weyl superfield and $K_\text{eff}$ and $W_\text{eff}$ are the effective K\"ahler potential and superpotential respectively. The latter is completely nonperturbative~\cite{Grisaru:1979wc}, while the former receives both perturbative and nonperturbative contributions.
The Weyl superfield $G$, capturing the induced coupling of the effective theory with gravity, carries left-handed spinor indices and its first two components are the self-dual gravitino field strength and the self-dual Riemann tensor respectively.

While the effective K\"ahler potential is largely unconstrained, the effective superpotential is stringently constrained by symmetry and holomorphy arguments and instanton calculations~\cite{Seiberg:1994bz, Intriligator:1994jr, Affleck:1983vc}.
Refs.~\cite{Dijkgraaf:2002dh, Dijkgraaf:2002fc} argued that the effective superpotential has the form
\begin{equation}
W_\text{eff} = \frac{\partial {\cal F}_0}{\partial S} 
            +  \frac{\partial^2 {\cal F}_0}{\partial S^2} G^{\alpha\beta\gamma}G_{\alpha\beta\gamma}  
            + {\cal O} \big((G^{\alpha\beta\gamma}G_{\alpha\beta\gamma})^2\big)\,,
\label{Weff}            
\end{equation}
where ${\cal F}_0$ is a holomorphic function of the glueball superfield $S$.\footnote{We do not include a detailed expression of the higher-order terms because they contribute only to higher-point gravitational amplitudes, so we do not need them.} Given a tree-level superpotential $W_\text{tree}$, the nonperturbative effective potential $W_{\text{eff}}$ can be computed algorithmically, including its gravitational couplings, either via a symmetry and holomorphy analysis \cite{Seiberg:1994bz, Intriligator:1994jr}, or through matrix-model methods \cite{Dijkgraaf:2002dh, Dijkgraaf:2002fc, Bena:2002kw, Dijkgraaf:2002xd, Dijkgraaf:2003sk, Cachazo:2002ry, Cachazo:2002zk, Cachazo:2003yc}. For our purpose here we do not need its detailed form.

In the latter approach most terms in $W_{\text{eff}}$ are evaluated perturbatively, and the glueball superfield $S$ enters initially in the form of the gauge invariant bilinear $S\propto {\rm Tr}[W^\alpha W_\alpha]$, where $W$ is the vector superfield, whose lowest component is the gluino. The nonperturbative nature of the superpotential comes from interpreting this superfield as a fundamental field, and from the inclusion in ${\cal F}_0$ of the Veneziano-Yankielovicz superpotential~\cite{Veneziano:1982ah}, which accounts for the chiral anomaly.

The critical point of the $G$-independent part of the effective superpotential, which is a solution of $\partial^2 {\cal F}_0/\partial S^2=0$, fixes a vacuum expectation value, $S=S_*$, of the glueball superfield. This breaks chiral symmetry and determines the non-normalized mass of this superfield as ${\tilde m}_S = \partial^3 {\cal F}_0/\partial S^3|_{S=S_*}$.\footnote{The mass $m_S$ depends on the details of the K\"ahler potential. On dimensional and holomorphy grounds one may expect that up to numerical factors, $m_S \propto S_*^{4/3} {\tilde m}_S$.} It also implies that among the terms with two Weyl-superfields there exists a linear coupling to $(S-S_*)$, $ m_S (S-S_*) G^{\alpha\beta\gamma}G_{\alpha\beta\gamma}$, which is also proportional to the mass of these fields. Evaluating the integral over Grassmann variables in Eq.~\eqref{Weff}, integrating out the the auxiliary fields and normalizing the quadratic term for 
$\varphi$ leads to
\begin{align}
\int d^2 \theta W_{\text{eff}} +h.c. = |{m}_S|^2 \bar\varphi \varphi + 
M\, \varphi\, R_{\alpha\beta\gamma\delta}R^{\alpha\beta\gamma\delta}
+
M^*\, {\bar \varphi}\, R_{\dot\alpha\dot\beta\dot\gamma\dot\delta}R^{\dot\alpha\dot\beta\dot\gamma\dot\delta}
+\text{fermions} \ ,
\label{Weffcomponents}
\end{align}
where $M= {\tilde m}_S/(\partial_S\partial_{\bar S} K_\text{eff}|_{S=S_*})^{1/2}$,  $\varphi = (S-S_*)|_{\theta = 0}$ is the scalar in the glueball superfield and $R^{\alpha\beta\gamma\delta}$ and $R^{\dot\alpha\dot\beta\dot\gamma\dot\delta}$ are
the self-dual and anti-self-dual parts of the Riemann tensor, 
containing the negative- and positive-helicity gravitons respectively. 
Thus, the nonperturbative superpotential couples the fluctuation $\varphi$ of the glueball scalar around its expectation value $S_*$ with two gravitons of the 
same helicity,
\begin{align}
\vcenter{\hbox{\scalebox{1}{\threePtAmpSGG}}} \hspace{-.4cm}\sim {\tilde m}_S
\quad
\rightarrow 
\quad
\vcenter{\hbox{\scalebox{1}{\threePtAmpggPhi}}}.
\end{align}
Setting aside the exchange of gravitons, the leading-order terms in the four-graviton amplitude come from tree-level exchange of $\varphi$. There are three possible contributions: two vertices from the superpotential, two vertices from the K\"ahler potential, and a contact term from the K\"ahler potential. 
Since extracting gravitons out of the $G$ superfield requires only one 
Grassmann derivative, it follows that the third field in a vertex with two 
gravitons is acted upon by two Grassmann derivatives, so it must be the 
auxiliary field of the glueball superfield. Thus, the K\"ahler potential can contribute only contact terms at tree level.
We graphically denote superpotential contributions by circles and K\"ahler potential contributions by boxes. The corresponding graphs are:
\begin{align}
\vcenter{\hbox{\scalebox{.8}{\exchangeSuperSuper}}}\,,
\hskip 1 cm 
\vcenter{\hbox{\scalebox{.8}{\exchangeKahlerKahler}}}\,,
\hskip 1 cm 
\vcenter{\hbox{\scalebox{.8}{\contactKahler}}} \,.
\end{align}
We first focus on the superpotential contributions and argue later that under certain circumstances the K\"ahler potential contributions do not affect the conclusions.

Four-graviton tree-level diagrams with vertices from the superpotential are very simple. Eq.~\eqref{Weffcomponents} implies that, for a fixed graviton helicity configuration, the amplitude receives contributions from a single exchange diagram.
For example, if gravitons 1 and 2 have negative helicity and gravitons 3 and 4 have positive helicity, the only diagram that contributes is: 
\begin{align}
 \mathcal{M}(1^-,2^-,3^+,4^+) \quad \sim \vcenter{\hbox{\scalebox{1}{\hmhmhphp}}}.  
\end{align}
This diagram depends on the scalar-field propagator, which in turn depends on the K\"ahler potential.\footnote{A dimensional-analysis-based suggestion was put forth in Ref.~\cite{Veneziano:1982ah}, $K(S)\propto ({\bar S} S)^{1/3}$.}
The  K\"all\'en-Lehmann representation of massive two-point functions identifies  this propagator as some function $f$ of the momentum with a simple pole at $p^2 = m_S^2$ where $m_S$ is the physical mass of the field $\varphi$. In particular, this implies that it has a regular expansion around vanishing momentum, i.e.
\begin{equation}
f(s) = \sum_{n\ge 0} a_n s^n \ .
\end{equation}
Then, the amplitude ${\cal M}(1^-, 2^-, 3^+, 4^+)$ is given by
\begin{align}
{\cal M}(1^-, 2^-, 3^+, 4^+)&= {\ab{12}^4[34]^4}\, f(s)  \ ,
\end{align}
with other helicity configurations, ${\cal M}(1^-, 2^+, 3^+, 4^-)$ and ${\cal M}(1^-, 2^+, 3^-, 4^+)$, obtained by relabeling. This implies that
\begin{align} 
\frac{a_{n,j}}{a_{n,0}} = 0\,, \ \text{ for all } j>0\,,
\label{ratio}
\end{align}
which puts the nonperturbative contributions from the superpotential on the small islands of physical EFTs and the low-spin-dominance lines in $D=4$ discussed in Ref.~\cite{Bern:2021ppb}.

Consider now briefly the consequences of including K\"ahler-potential contributions to leading-order gravitational amplitudes. As we saw earlier, these contributions can come only from contact terms which arise either from terms of the type $\int d^4\theta f(S, {\bar S}) G^2{\bar G}^2$ in $K_\text{eff}$ in Eq.~\eqref{Lagrangian}, or from terms with two Weyl multiplets in $K_\text{eff}$ upon integrating out the auxiliary field in the glueball superfield.

Nonperturbative contributions to $K_\text{eff}$ necessarily depend on the confinement scale $\Lambda$; moreover, if they are dependent on momenta, then dimensional analysis suggests that they depend on $p^2/\Lambda^2$ which, according to our initial assumption, is negligible at the energy scale $p^2\ll \Lambda^2$ that we are focusing on.
Thus, the contributions of such K\"ahler potential terms to amplitudes are essentially constant (up to the helicity-dependent factor), and therefore they affect only a limited number of $a_{n,j}$ coefficients, perhaps only $a_{0,0}$, leading to effectively no changes to Eq.~\eqref{ratio}.

One may wonder if the expansion around the nonvanishing  expectation value for the glueball superfield $S$ may enhance these terms suppressed by the ratio $p^2/\Lambda^2$. The expectation of a smooth limit in which the high-energy theory is trivial (i.e. that the tree-level superpotential is zero) suggests that the momentum dependence cannot be enhanced for small values of the parameters of the high-energy theory. 
While this argument suggests that the nonperturbative momentum dependence could be enhanced at large values of these parameters, naively of the order of $\Lambda^2/p^2$, 
there remains a comfortable range of parameters of the high-energy theory for which inclusion of the purely nonperturbative terms in the K\"ahler potential does not affect Eq.~\eqref{ratio}. 

Extension of this discussion to the perturbative and mixed part of the K\"ahler potential is difficult because of their detailed dependence on the parameters of the high-energy theory. Using however the identity 
\begin{align}
    x\le \frac{a^{(1)}_{n,j}}{a^{(1)}_{n,0}}, \, \frac{a^{(2)}_{n,j}}{a^{(2)}_{n,0}} \le y
    \quad \Longrightarrow \quad
    x\le \frac{a^{(1)}_{n,j}+a^{(2)}_{n,j}}{a^{(1)}_{n,0}+a^{(2)}_{n,0}} \le y \ ,
\end{align}
if all $a_{n,j}>0$ together with the results of Ref.~\cite{Bern:2021ppb} and those discussed in later sections, we expect that the complete nonperturbative four-graviton amplitude in the class of theories discussed here belongs to the EFT island for $p^2/\Lambda^2\ll 1$.
Further study and explicit calculations are necessary to fully settle this issue and to extend our analysis to scales $p^2 \lesssim \Lambda^2$ in which all nonperturbative contributions to the Wilsonian effective action become important.

\section{Tree-level graviton amplitudes in string theory} 
\label{sec:string_data}

Having analyzed the first nonperturbative data for four-dimensional graviton scattering via supersymmetric arguments, we now move on to more traditional perturbative amplitudes, although beyond the commonly considered four-dimensional setup. We expect that this data will provide useful guidance for any attempts to place dispersive bounds on graviton scattering in higher dimensions. 

To this end, we collect the available tree-level results for the scattering of four external gravitons in superstring (ss), heterotic-string (hs), and bosonic-string (bs) theory. These closed-string amplitudes are determined by Kawai-Lewellen-Tye (KLT) relations \cite{Kawai:1985xq} in terms of open-string ones. We take the mass of the first excited string level to be $m^2_s = 4/\alpha'$ for all string theories and express the string-theory amplitudes in terms of gauge-invariant tensor structures $\mathcal{T}$. We need two such structures,
\begin{align}
\begin{split}
   \mathcal{T}_{\text{sYM}} & = -
   s\, t\, (\varepsilon_1\cdot\varepsilon_3) (\varepsilon_2\cdot \varepsilon_2) + \cdots \,,
   \\
   \mathcal{T}_{\text{bos}} & = -
   \frac{s\, u}{(1+ \alpha' t/4)} (\varepsilon_1\cdot \varepsilon_4)(\varepsilon_2\cdot\varepsilon_3) + \cdots\,,
\end{split}   
\end{align}
which are normalized to have mass-dimension four. The full expression for these structures is provided in the ancillary file. We may now write the four-graviton amplitudes in a form that uniformly applies to the three string theories,
\begin{align}
\begin{split}
\hspace{-.4cm}
    \mathcal{M}^{(\beta)} = {-}\left(\frac{\kappa}{2}\right)^2\!\! \left(\frac{\alpha'}{4} \right)^3 \!\!
    \mathcal{T}_\beta\,
    \frac{
    \Gamma\!\left[{-}\frac{\alpha' s}{4}\right]
    \Gamma\!\left[{-}\frac{\alpha' t}{4}\right]
    \Gamma\!\left[{-}\frac{\alpha' u}{4}\right]
    }
    {
    \Gamma\!\left[1{+}\frac{\alpha' s}{4}\right]
    \Gamma\!\left[1{+}\frac{\alpha' t}{4}\right]
    \Gamma\!\left[1{+}\frac{\alpha' u}{4}\right]
    },
\hspace{-.4cm}    
\end{split}
\end{align}
where $\beta \in \{ \text{ss, hs, bs} \}$ and
\begin{equation}
   \mathcal{T}_\text{ss} = \mathcal{T}^2_\text{sYM}\,, 
   \qquad
   \mathcal{T}_\text{hs} = \mathcal{T}_\text{sYM}\, \mathcal{T}_\text{bos}\,,
   \qquad
   \mathcal{T}_\text{bs} = \mathcal{T}^2_\text{bos}\,.
\end{equation}
Newton's constant $G$ is related to $\kappa$ via $\kappa^2 = 32\pi G$. As a consistency check, by specializing the polarization tensors to describe four-dimensional helicity states we recover the results collected in Appendix B of Ref.~\cite{Bern:2021ppb}.

The bosonic string-theory amplitude contains the tachyon exchange. Following Ref.~\cite{Bern:2021ppb}, we define a modified bosonic string-theory amplitude by subtracting this exchange and present our data within this definition. In four dimensions, this amplitude is consistent with the generic bounds and hence we expect the same to be true in higher dimensions as well. Let us, however, note that none of the conclusions drawn in the present paper change if we chose to drop the bosonic string-theory amplitude, given the undesired appearance of the tachyon.

We may also identify the massless exchanges in all three sting theories. Specifically, the superstring amplitude only contains the minimal-coupling graviton exchange. The heterotic string-theory amplitude contains both the minimal-coupling and Gauss-Bonnet graviton exchanges, as well as the dilaton exchange. Finally, the bosonic string-theory amplitude contains all the heterotic string-theory amplitude exchanges as well as those from an $R^3$-type coupling. Subtracting these exchange contributions is possible but not necessary in order for the amplitudes to be physical. We do not perform such a subtraction in the present work.

Furthermore, we point out that there is an inherent ambiguity in separating the non-analytic from the analytic part (or equivalently the non-local from the local part) of the amplitude. This ambiguity is reflecting part of the freedom in writing a Lagrangian. When there are no $R^3$-type contributions, there is no such ambiguity. In these cases, a simple powercounting argument shows that dilaton and Gauss-Bonnet contributions do not mix with $D^nR^4$-type (contact) operators. In contrast, when $R^3$-type operators are present, terms of the form $R^2 \times R^3$ and $R^3 \times R^3$ contribute at the same order as $R^4$-type and $D^2R^4$-type operators respectively. Hence, in order to obtain the coefficients of the latter, one needs to calculate the low-energy amplitudes starting from a Lagrangian (or define in a different way a scheme for separating the non-analytic from the analytic part of the amplitude). For the complete analysis of the inherent ambiguity in mapping amplitude coefficients to Lagrangian coefficients one has to also include massless loop effects~\cite{Bellazzini:2020cot,Bellazzini:2021oaj}. In this work we neglect massless loop effects, and plot coefficients of $D^nR^4$-type operators with $n>2$ which do not mix with the massless exchange contributions.

\section{Loop-level graviton amplitudes in quantum field theory} 
\label{sec:FT_data}

Besides the tree-level string theory amplitudes discussed in the previous section, we are also interested in field-theory models where we allow massive states circulating in the loop as a $D$-dimensional extension of the four-dimensional analysis of Ref.~\cite{Bern:2021ppb}. Such models should not be viewed as proper UV completions, but should instead be viewed as intermediate-energy theories.  Because they satisfy all input assumptions used to derive bounds on the low-energy EFTs they provide useful guidance on where physically sensible theories live. Here we opt to construct the full amplitudes for a simple reason:  In the derivation of EFT bounds one needs the Regge behavior to determine the validity of bounds which rely on knowing (or assuming) the high-every behavior. With the exact amplitudes in hand it is straightforward to extract the high-energy behavior. A side benefit in having the full one-loop amplitudes computed is that they may be useful for purposes other that studying low-energy EFTs.

\subsection{Maximally supersymmetric massive matter in the loop}
\label{massiveKKmatter}

The simplest loop-level data we can consider originates from a massive deformation of the maximally supersymmetric gravity amplitude at one-loop, where we give a common mass to the supermultiplet circulating in the loop. This is accomplished by taking the original massless one-loop four-point amplitude written in terms of scalar box integrals \cite{Green:1982sw},
\begin{align}
\hspace{-.3cm}
    \mathcal{M}^{(\mathcal{N}{=}8)} = \left(\frac{\kappa}{2}\right)^4 \mathcal{T}_\text{ss} \left[
    I^{(D)}_{{\rm box}} (s,t) 
    + I^{(D)}_{{\rm box}} (s,u)
    + I^{(D)}_{{\rm box}} (u,t)
    \right]\,.
\end{align}
and replacing the loop propagators by massive ones. This may be interpreted as dimensionally-reducing a higher-dimensional maximal supergravity and integrating out a Kaluza-Klein mode whose mass is the extra-dimensional momentum. The polarization information is encoded in the tree-level superstring tensor $\mathcal{T}_\text{ss}$ and the integrals are defined e.g.~as
\begin{align}
    I^{(D)}_{{\rm box}} (s,t) =
    \int \frac{d^D\ell}{(2\pi)^D}
    \prod_{n=0}^3   \frac{1}{\left(\ell+\sum_{i=1}^n p_i\right)^2-m^2}\,.
\end{align}
The UV divergences that are present for $D\ge 8$ are straightforward to extract. In fact, the UV expansion of the amplitude and its large-mass expansion are intimately related: both are equivalent to an expansion around small external momenta yielding massive tadpoles, as can be easily seen from the interpretation of the mass as Kaluza-Klein momentum. We will return shortly to the large-mass expansion.

\subsection{Non-supersymmetric massive matter in the loop} 

We obtained the four-graviton amplitude with a maximally supersymmetric massive multiplet circulating in the loop by simply replacing formerly massless propagators 
by massive ones in the known representation of the one-loop amplitude~\cite{Green:1982sw}.
We are also interested in $D$-dimensional one-loop amplitudes with massive scalars and massive spin-1 fields running in the loop. We do not explicitly consider massive fermions, gravitinos, or massive spin-2 states here for the sake of brevity even though there is no conceptual problem to construct these amplitudes as well. The workflow of our construction uses a number of modern scattering-amplitude methods.

We start by constructing the amplitude's loop integrand via generalized unitarity \cite{Bern:1994zx,Bern:1994cg,Britto:2004nc} from the knowledge of tree-level amplitudes. We find it most convenient to directly match two-particle cuts
\begin{align}
\label{eq:bubble_cut}
\vcenter{\hbox{\scalebox{1}{\bubCut}}} \,,
\end{align}
where the external states are gravitons (denoted by wiggly lines) and the massive states propagating inside the loop are generically denoted by a solid line (which can represent either massive spin-0 or spin-1 states). Since we are interested in expressions that are valid in arbitrary spacetime dimension $D$, we utilize tree-level gravitational Compton amplitudes (grey `blobs' in (\ref{eq:bubble_cut})) in terms of formal (traceless symmetric) polarization tensors $\varepsilon^{\mu\nu}_i= \varepsilon^\mu_i \varepsilon^\nu_i$ of the gravitons that can be found e.g.~in Ref.~\cite{Bern:2020buy}.
The unitarity cuts involve a sum over physical states. For the massive spin-1 circulating in the loop, the relevant states are selected out by inserting the physical state projector $ \Pi^{\mu \nu}(p,m)=\eta^{\mu\nu} - \frac{p^\mu p^\nu}{m^2}$ for each cut leg. From the resulting expressions, we extract the kinematic numerators of the cubic diagrams
\begin{align}
\vcenter{\hbox{\scalebox{1}{\boxInt}}}\,,
\hskip 1 cm
\vcenter{\hbox{\scalebox{1}{\triInt}}}\,,
\hskip 1 cm 
\vcenter{\hbox{\scalebox{1}{\bubInt}}}
\,,
\end{align}
which are functions of the following Lorentz invariants: $\varepsilon_i\cdot p_j$, $s,t$, $p_i\cdot \ell$, and $\varepsilon_i\cdot \ell$. For the massive spin-1 exchange, it is convenient to split the numerators into a `spin-0' part and a `spin-1' remainder akin to the supersymmetric decomposition of the four-dimensional amplitudes in Ref.~\cite{Bern:2021ppb}:
\begin{align}
\label{eq:spin_1_decomposition}
    N^{\text{spin}-1} = \overline{N}^{\text{spin}-1} + (D_s-1)N^{\text{spin}-0}\, ,
\end{align}
where $D_s = \eta^\mu_\mu$ is the state-counting parameter. This separation effectively eliminates the terms of highest degree in the loop momentum $\ell$ from $N^{\text{spin}-1}$.

At one loop, all contractions between the loop momentum $\ell$ and the external momenta $p_i$ can be written in terms of inverse propagators, see e.g.~\cite{Smirnov:2012gma}, but contractions of the loop-momentum with polarization vectors $\varepsilon_i$ require the evaluation of tensor integrals. Such integrals are well known from e.g.~Refs.~\cite{Tarasov:1996bz,Tarasov:1997kx} where one converts tensor integrals into dimension-shifted \cite{Tarasov:1996br,Lee:2009dh,Lee:2010wea} scalar integrals, see also Ref.~\cite{Anastasiou:2005cb}. The relevant dimension shifts can be derived algorithmically with the help of integration-by-parts relations \cite{Tkachov:1981wb,Chetyrkin:1981qh,Laporta:1996mq,Laporta:2001dd}, implemented in modern computer codes such as \texttt{FIRE} \cite{Smirnov:2008iw,Smirnov:2019qkx}.

We find it convenient to organize the resulting amplitudes in a special basis of scalar integrals where all integral coefficients are independent of the mass of the state in the loop and of the spacetime dimension $D$. This closely follows the discussion of the four-dimensional amplitude construction of Ref.~\cite{Bern:2021ppb} and can be likewise achieved by judiciously using dimension-shifting relations~\cite{Tarasov:1996br,Lee:2009dh,Lee:2010wea}. 

The attentive reader might have noticed that our amplitudes construction via the two-particle cut in Eq.~(\ref{eq:bubble_cut}) is not yet complete as we could be  missing contributions from bubble integrals with a single massless line on one side (sometimes referred to a `snail integrals'), or from tadpole integrals. These contributions are determined along the lines of Ref.~\cite{Bern:1995db} by demanding that the amplitudes considered here remain IR-finite in the massless limit and by requiring that the UV divergences are mass-independent.\footnote{The latter requirement is equivalent to demanding that the corresponding counterterm has a local expression in terms of Riemann tensors.} At the end of our construction, we find that all one-loop four-point amplitudes due to massive spin-$\rm S$ exchange considered in this work can be expressed in terms of 42 independent (possibly dimension shifted) integrals that are multiplied by gauge-invariant tensors involving only the Mandelstam invariants and contractions between polarization vectors and external momenta,
\begin{align}
\label{eq:1loop_amps_abstract}
    \mathcal{M}^{({\rm S})} = \sum_{k} \mathcal{T}^{({\rm S})}_k(\varepsilon,p)
    \ I_k(s,t,D,m^2)\,.
\end{align}
We include the explicit expressions of the amplitudes in the representation of (\ref{eq:1loop_amps_abstract}) in computer readable form with explicit rules for the $\mathcal{T}^{({\rm S})}_k$ in terms of Lorentz products of the external data.  

In view of our goal to provide explicit expressions for the low-energy gravitational EFT amplitudes once the massive state is integrated out, it is highly advantageous to organize the amplitudes as in Eq.~(\ref{eq:1loop_amps_abstract}). Indeed, since only the scalar (dimension-shifted) integrals, $I_k(s,t,D,m^2)$, depend on the mass, we only need to find their large-mass expansion. This can be achieved by dimension shifting all integrals back to $D=4$ and using the known polylogarithmic expressions, conveniently collected in Ref.~\cite{Ellis:2007qk}, and expanding them for $m^2\gg |s|,|t|$. Alternatively, the same expressions can be obtained by expanding the integrand along the lines of the `method of regions' \cite{Beneke:1997zp} for $|\ell| \sim m \ll |p_i|$, leading effectively to tadpole integrals which are known exactly in $D$. We find the latter method computationally much more efficient. We provide the expansion of all one-loop integrals that appear in (\ref{eq:1loop_amps_abstract}) to sufficiently high order in $1/m^2$ in the attached ancillary file.

As a further consistency check on our computation, we can evaluate our $D$-dimensional scattering amplitudes for $D=4-2\epsilon$ and for specific choices of polarization vectors $\varepsilon^\mu_i$ that correspond to four-dimensional helicity states, 
thereby reproducing all earlier results from Ref.~\cite{Bern:2021ppb}, e.g.
\begin{align}
\label{eq_d4_allPlus_data}
\hspace{-.5cm}
    \mathcal{M}^{(0)}(1_{4^+}2_{4^+}3_{4^+}4_{4^+}) = 
      \frac{stu}{504 m^2} 
    + \frac{(s^2 +st+t^2)^2}{3780 m^4} 
    + \ldots \,,
\hspace{-.5cm}    
\end{align}
where $1_{4^+}$ refers to the polarization vector $\varepsilon_{1,4^+}^\mu$, etc. (See our conventions in Eq.~(\ref{eq:pol_def})). 

Using our scattering amplitudes we may now probe the richer space of states of the graviton in $D>4$. We consider examples of elastic amplitudes of the form $\mathcal{M}(1_{a^-}2_{b^-}3_{b^+}4_{a^+})$ for various values of $a$ and $b$.

As mentioned above, generally, in order to ensure the validity of EFT bounds it is crucial to know the behavior of our amplitudes in the Regge limit, $|s|\gg-t,\,|s|\gg m^2$. Since our field theory data are only stand-in models for a intermediate UV completion of gravitational scattering, their behavior is generically worse than expected from quantum Regge bounds~\cite{Maldacena:2015waa,Chandorkar:2021viw,Haring:2022cyf}. The exact behavior of the amplitudes in the Regge limit depends on the graviton polarizations, but using the explicit expressions for the amplitudes we have checked that the worst behavior is saturated by the spin-2 exchange (inside the $\mathcal{N}=8$ amplitude) with a scaling (in even spacetime dimension $D$) of the form
\begin{align}
    \mathcal{M} \sim s^{D-1}\,,
\end{align}
which recovers the $s^3$ behavior in $D=4$ explored in Ref.~\cite{Bern:2021ppb}.

Similarly to the string-theory cases, the non-supersymmetric examples we consider here contain massless exchanges that would need to be taken into account in mapping the amplitude coefficients to Lagrangian coefficients.

\section{Data summary and plots} 
\label{sec:plots}

Having discussed the relevant computations of the new explicit models of UV completions, we now proceed to explore the associated values of the low-energy couplings in the large-mass expansion of the amplitudes. Our analysis supports the notion that physical theories lie on small islands. Interestingly, if we choose the external states to be $4_\pm$ (which may be thought of as polarization tensors restricted to a four-dimensional subspace), we find that the projective data points remain on the same four-dimensional islands independent of the spacetime dimension.

\subsection{Sample data in $D=4$} 

To confirm our $D$-dimensional setup, we first reproduce the data points in Fig.~10 of Ref.~\cite{Bern:2021ppb} by specializing to $D=4$. To do so we select four-dimensional helicity states for the external gravitons corresponding to  $\mathcal{M}(1_{4^-}2_{4^-}3_{4^+}4_{4^+}) = s^4 f(t,u)$ (see Eq.~(\ref{eq:pol_def}) for our polarization conventions), where $f(t,u)$ admits the low-energy expansion 
\begin{align}
    \label{eq:4d_low_energy_expansion}
    f(t,u) = \sum_{k\geq q \geq 0} a_{k,q}\,  s^{k-q} t^q\,.
\end{align}
Additionally, as explained in section~\ref{sec:np_data}, we add a new data point for our nonperturbative results for the four-graviton scattering generated from the effective superpotential of a $\mathcal{N}=1$ matter-coupled supersymmetric gauge theory further coupled to gravity. In this special case, $f$ is a function of $s$ only so that the sum truncates and the only nonzero coefficients are the $a_{k,0}$. 

\begin{figure}[ht!]
\centering
    \includegraphics[scale=1]{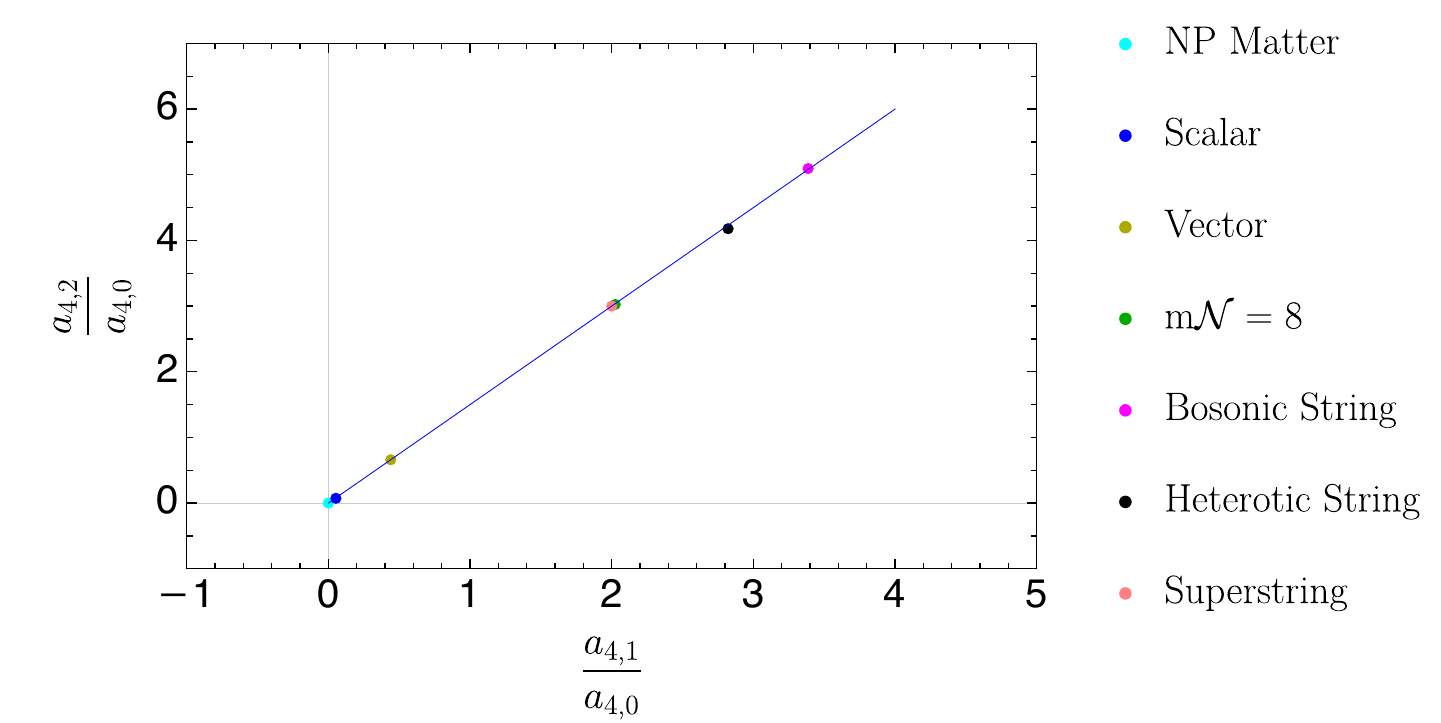}
\vspace{-.3cm}
\caption{
\label{fig:d4_mmpp_k4}
The $D=4$ EFT data for various models for ${a_{4,1}/ a_{4,0}}$ and ${a_{4,2}/a_{4,0}}$. A line with slope 3/2 is added. The data points do not land perfectly on this line.}
\end{figure}

In Fig.~\ref{fig:d4_mmpp_k4} and the following, our labeling conventions are as follows: NP Matter $\equiv$ nonperturbative matter-coupled $\mathcal{N}=1$ supersymmetric gauge theory, Scalar $\equiv$ massive spin-0 running in the loop, Vector $\equiv$ massive spin-1 running in the loop, $\text{m}\mathcal{N}{=}8\equiv$ massive $\mathcal{N}=8$ supermultiplet in the loop. The new data points further emphasize the main observation of Ref.~\cite{Bern:2021ppb}, that explicit data lies on small `theory islands' in the space allowed by unitarity, causality, and crossing constraints (this larger space is not indicated in Fig.~\ref{fig:d4_mmpp_k4} and is the red-shaded region in Fig.~9 of Ref.~\cite{Bern:2021ppb}).

\subsection{Sample data in $D=6$} 

\begin{figure}[ht!]
\centering
\includegraphics[scale=.98]{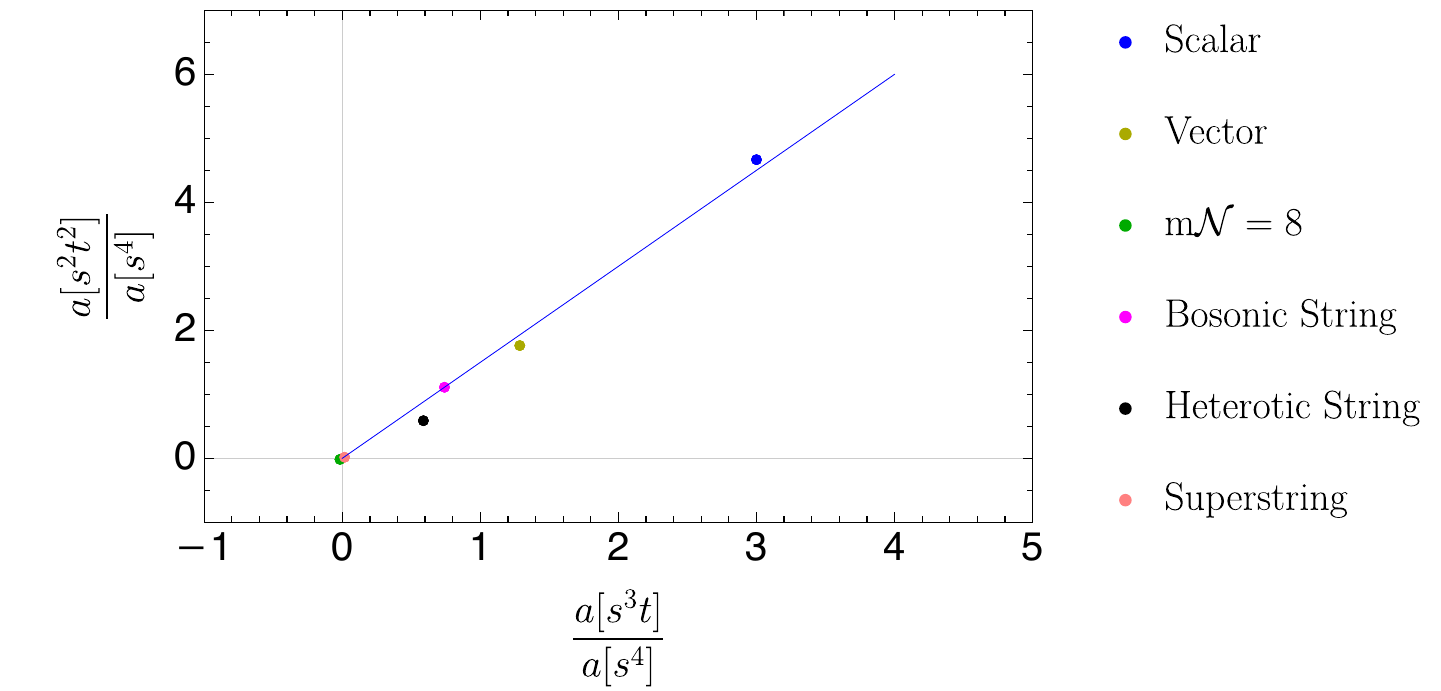}
\vspace{-.3cm}
\caption{
\label{fig:d6_m6m6p6p6_k0}
The $D=6$ EFT data for various models for $a[s^3 t ]/ a[s^4]$ and $a[s^2t^2]/a[s^4]$. A line with slope $3/2$ is added to guide the eye. Here $a[x]$ stands for the coefficient of the monomial $x$ in the Taylor expansion of the amplitude.
}
\end{figure}

To showcase some features of our amplitude data, we generate similar data plots for graviton polarizations that are outside the four-dimensional helicity setup. To this end, we first consider scattering of gravitons in $D=6$ with the polarization choice $\mathcal{M}(1_{6^-}2_{6^-}3_{6^+}4_{6^+})$ where all polarizations are outside the four-dimensional subspace. (See Eq.~(\ref{eq:pol_def}) for our conventions of the graviton polarization states.) We consider the coefficients of the $1/m^2$ terms in Fig.~\ref{fig:d6_m6m6p6p6_k0} which are polynomials in $s,t$ of degree four. The particular helicity choice renders the amplitude $t\leftrightarrow u$ symmetric and we denote the amplitude coefficients by their corresponding monomial in the Mandelstams. Unlike for the $D=4$ examples considered previously, it is not always possible to factor out some overall powers of Mandelstams.

It is fascinating to observe that, similarly to four-dimensional theories, the data points of the various models lie on an almost straight line and their spread from the line is much smaller than the extend of the line itself, giving us a concrete first example of small theory islands beyond $D = 4$.

\begin{figure}[ht!]
\centering
    \includegraphics[scale=1]{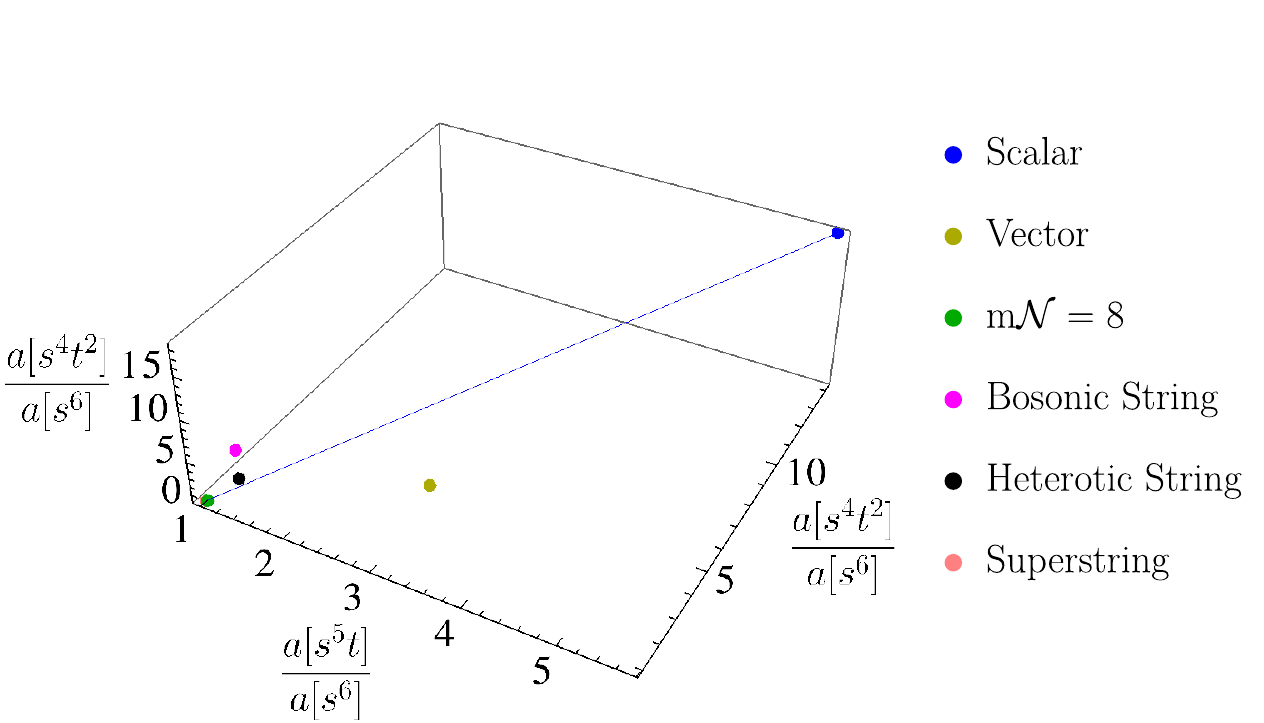}
\vspace{-.3cm}    
\caption{
\label{fig:d6_m6m6p6p6_k2}
The $D=6$ EFT data for various models for the ratios of low-energy amplitude coefficients $a[s^5 t]/ a[s^6]$, $a[s^4t^2]/a[s^6]$, and $a[s^3t^3]/a[s^6]$. We add a straight line to guide the eye. As illustrated in Fig.~\ref{fig:d6_m6m6p6p6_k2_rotated} from a different viewpoint the data essentially lie in a plane.
}
\end{figure}

\begin{figure}[ht!]
\centering
    \includegraphics[scale=.98]{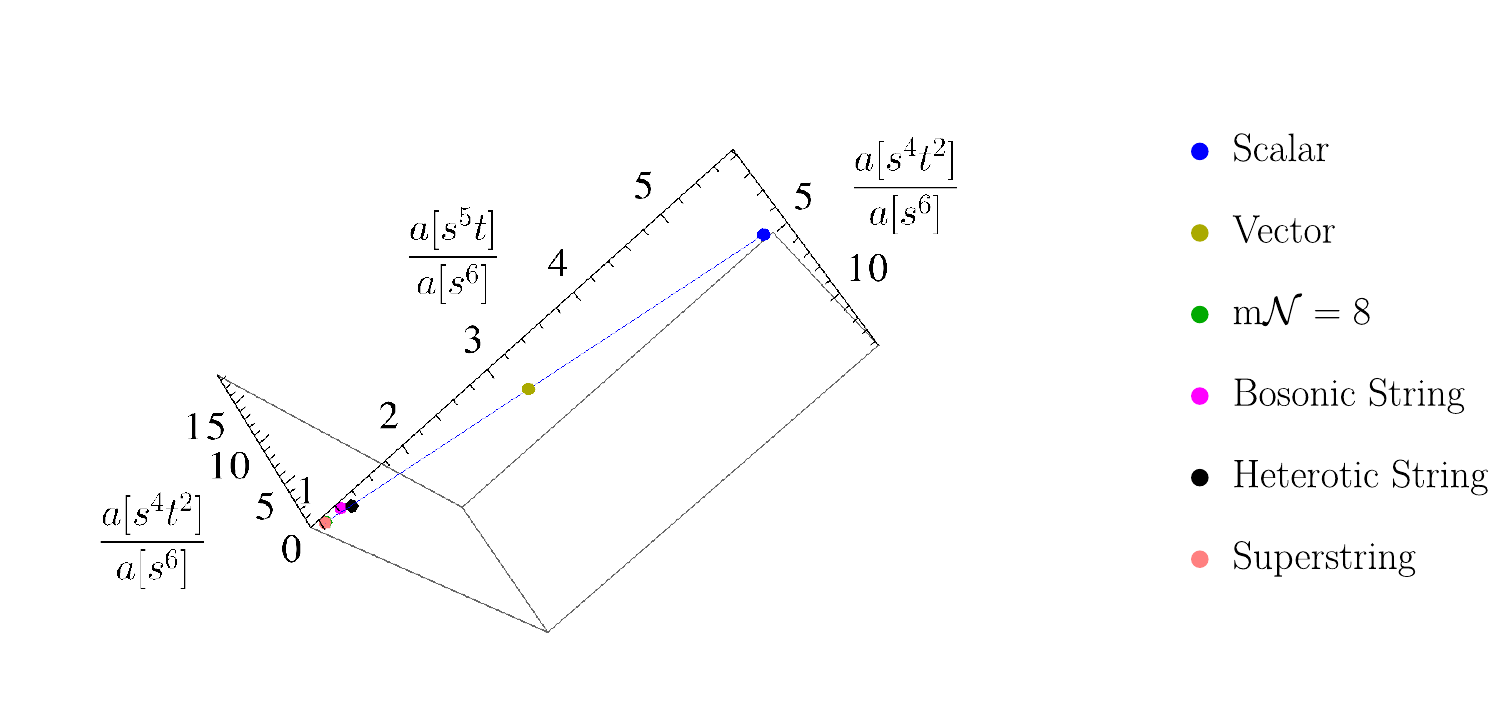}
\vspace{-.3cm}    
\caption{
\label{fig:d6_m6m6p6p6_k2_rotated}
The same data points and line of Fig.~\ref{fig:d6_m6m6p6p6_k2} from a different viewpoint that demonstrates that the data essentially lie in a plane.
}
\end{figure}

We should, however, stress that not all extra-dimensional data falls on such perfect lines. To see this, let us investigate a three-dimensional section of the coefficient space at mass-level $1/m^{6}$ where we have degree-6 polynomials in $s$ and $t$. We summarize our results in Fig.~\ref{fig:d6_m6m6p6p6_k2}.  As seen in a rotated viewpoint shown in Fig.~\ref{fig:d6_m6m6p6p6_k2_rotated}, this data essentially lies in a plane. 
While it appears that the virtual scalar EFT lies somewhat off a line formed 
by the other models, it is difficult to assess the broader significance of this departure vis-\`a-vis the parameter space allowed by causality, unitarity and crossing constraints, which is currently not known beyond $D=4$.

\subsection{Sample data in $D=10$} 

We proceed to analyze the data provided by perturbative calculations in $D=10$, 
which is an interesting dimension from a superstring perspective. We observe that, as in lower dimensions, the ratios of four-graviton amplitude coefficients again lie on a remarkably thin island.  As mentioned in the introduction, this theoretical data should provide crucial guidance for future dispersive analyses analogous to those carried out in $D=4$. 
Here, we only plot one particular section through our data in $D=10$ which could eventually interplay with the search for string theory via the analysis of graviton scattering. For concreteness, we consider the amplitude $\mathcal{M}(1_{6^-}2_{10^-}3_{10^+}4_{6^+})$. 

\begin{figure}[ht!]
\centering
    \includegraphics[scale=1]{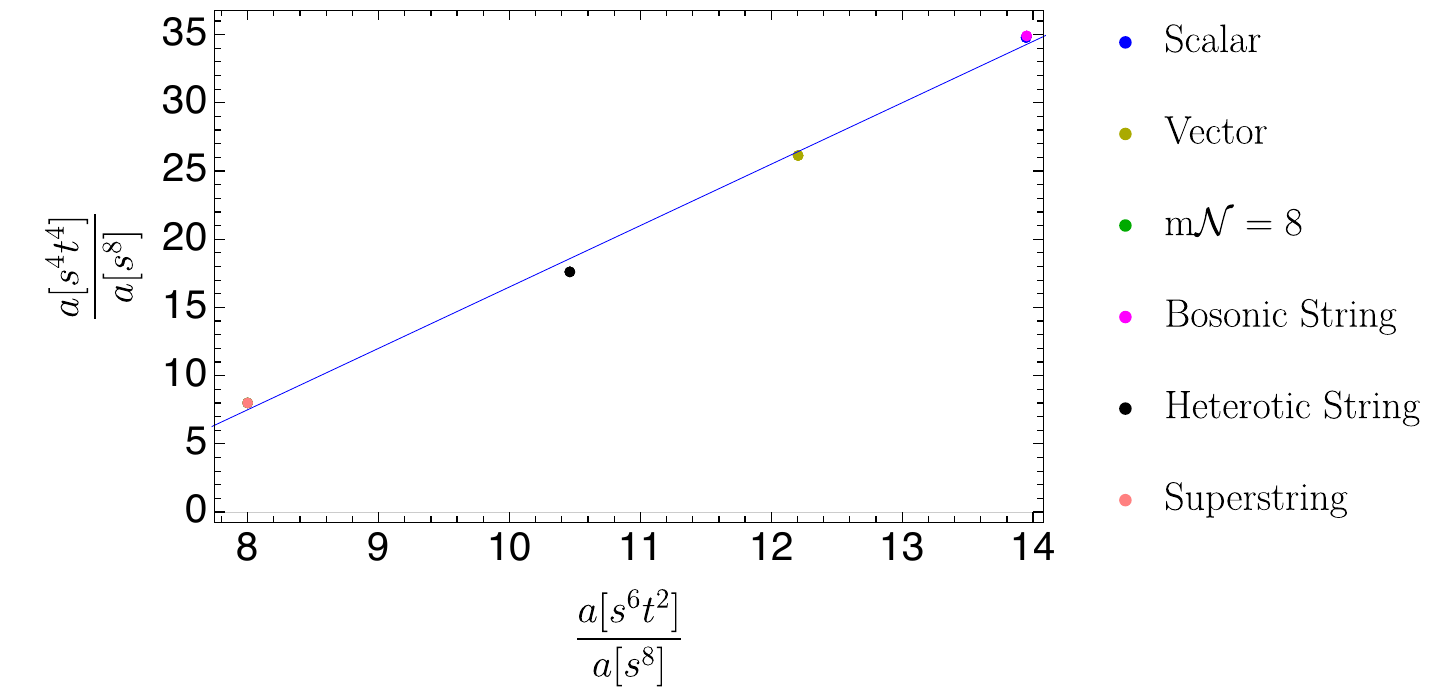}
\caption{The $D=10$ EFT data for various models for the ratios of low-energy amplitude coefficients $a[s^6 t^2 ]/ a[s^8]$ and $a[s^4t^4]/a[s^8]$. A line with slope $9/2$ is added to guide the eye.
}
\label{fig:d10_m6m10p10p6_k4}
\end{figure}

Taking the large-mass expansion and evaluating all tensor structures for the specified graviton-polarization choice we collect, for example, the expansion coefficients similar to the $k=4$ coefficients in $D=4$. Notably, in $D=10$ it no longer holds that we can factor out an overall helicity-dependent polynomial of the Mandelstam invariants. Therefore, the formerly $k=4$ amplitude coefficients are associated to honest degree-8 polynomials in $s,t$. In the particular example we discuss, even though we do not naively have crossing symmetry due to the polarization choice, the only nontrivial contractions of polarization vectors that survive are $\varepsilon_2\cdot \varepsilon_3$ and $\varepsilon_1\cdot \varepsilon_4$ which is left invariant under $2\leftrightarrow 3$ or $1\leftrightarrow 4$. At the level of the polynomials in Mandelstam invariants, this leaves three independent degrees of freedom that we can in general bound from unitarity, causality, crossing and Regge-behavior considerations. The three independent coefficients are associated to the $s^8$, $s^6 t^2$ and $s^4t^4$ terms respectively and their ratios are depicted in Fig.~\ref{fig:d10_m6m10p10p6_k4}.

The data lie on an almost straight line of approximate slope $9/2$. We leave to future work to establish the appropriate bounded regions of allowed parameter space from unitarity and causality constraints. However, the similarity between Figs.~\ref{fig:d4_mmpp_k4} and \ref{fig:d10_m6m10p10p6_k4} suggests that many of the interesting features of the four-dimensional theory islands survive in higher dimensions as well. 

\begin{figure}[ht!]
\centering
    \includegraphics[scale=1]{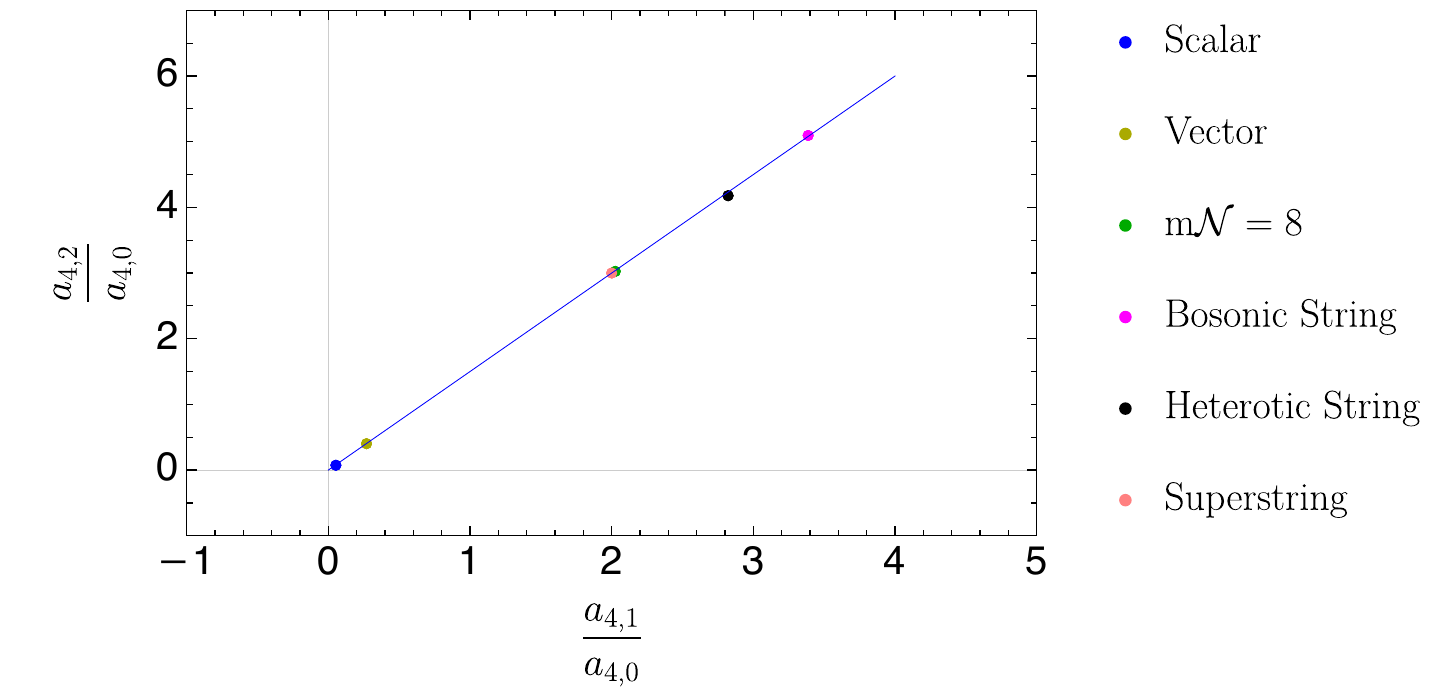}
\caption{The $D=10$ EFT data with four-dimensional external polarizations $4_\pm$. Here we follow the conventions of Eq.~(\ref{eq:4d_low_energy_expansion}). A line with slope 3/2 is added to guide the eye.
}
\label{fig:d10_m4m4p4p4_k4}
\end{figure}

Recently, the projective bounds between independent Wilson coefficients of the same mass-level $k$ have been generalized to extremely interesting bounds of Wilson coefficients against e.g. Newton's constant~\cite{Caron-Huot:2022ugt,Chiang:2022jep}. Here, we chose to plot projective data points due in part to the surprising observation that Fig.~\ref{fig:d4_mmpp_k4} in $D=4$ does not change significantly compared to Fig.~\ref{fig:d10_m4m4p4p4_k4} in $D=10$, where in both cases we evaluate the amplitudes for polarizations corresponding to four-dimensional helicity states. Specifically, only the ``Vector'' data point is $D$-dependent, and this dependence is solely due to the $D_s$ appearing in Eq.~(\ref{eq:spin_1_decomposition}). The net effect of this $D$-dependence is that the ``Vector'' data point is closer to the ``Scalar'' in Fig.~\ref{fig:d10_m4m4p4p4_k4} compared to Fig.~\ref{fig:d4_mmpp_k4}. While we do not spell out the details, this relative uniformity of the data across dimensions can be understood from the analysis of dimension-shifted scalar integrals. This is consistent with the intuition that extra-dimensional momenta in the loop can be thought of as a Kaluza-Klein mass which we effectively integrate over. This integral over the mass then drops out from the projective data.

\section{Conclusions and Outlook}
\label{sec:conclusions}

In this work we studied explicit perturbative and nonperturbative models for UV sensible four-graviton scattering amplitudes and their respective low-energy expansions. In particular, as a first example, we studied a nonperturbative~$\mathcal{N}=1$ supersymmetric gauge theory also coupled to gravity in four spacetime dimensions.  Moreover, in $D$ dimensions we considered tree-level four-graviton amplitudes in string theory and computed minimally-coupled one-loop four-graviton amplitudes with massive matter circulating in the loop. 

The crucial output is the low-energy expansion of these amplitudes. As in the four-dimensional EFTs considered in Ref.~\cite{Bern:2021ppb}, the low-energy coefficients of our data populate rather small theory islands compared to the naive expectation that consistent theories should fill out the full space of allowed low-energy couplings. Unlike the four-dimensional case where this feature was attributed to low-spin dominance, in general spacetime dimensions we do not currently have the same level of understanding. Nonetheless, our explicit analysis suggests that similar mechanisms are at work and it would be extremely interesting to further explore it by studying the partial-wave decomposition of our higher-dimensional amplitudes. We also anticipate that our explicit data will serve as a useful guide for any upcoming analysis of dispersive bounds on the low-energy Wilson coefficients. Further data can be found by exploiting the results of Ref.~\cite{Edison:2021ebi}, in which one-loop amplitudes in theories deformed by operators induced by integrating out massive string states are computed. Similarly to the discussion in section~\ref{massiveKKmatter}, the states circulating in the loop can be rendered massive through Kaluza-Klein reduction; then, the resulting amplitudes are interpreted as those of an EFT of compactified string theory valid at scales $p^2\sim m_\text{KK}^2\ll (\alpha')^{-2}$, in the same spirit as the discussion in section~\ref{sec:np_data}.

There are several further interesting directions to pursue. For one, beyond $D=4$, gravitational scattering amplitudes are no longer plagued by infrared singularities that hamper obtaining a number of four-dimensional bounds from various approaches and can show up in the form of IR-logarithms even beyond the forward limit bounds \cite{Caron-Huot:2022ugt}. Gravitational scattering in higher dimensions is also interesting in light of the recent attempts to find string theory~\cite{Guerrieri:2021ivu} from an S-matrix bootstrap point of view. It would be very interesting to narrow down the range of possible extensions of gravitational UV completions in 10 spacetime dimensions beyond string theory. One could also combine the scattering matrix for different graviton states in higher dimensions into a matrix, on which nontrivial bounds can be determined, analogous to the ones found for different helicity configurations in four dimensions~\cite{Bern:2021ppb}.

Additionally, the data presented here is primarily deduced from conventional gravitational scattering amplitudes, although we also presented a first example of data deduced from nonperturbative amplitudes. Recently, certain more exotic amplitudes with ``accumulation point spectra'' have gained some attention in e.g.~ Refs.~\cite{Figueroa:2022onw,Huang:2022mdb} and likewise in the model of Appendix D of Ref.~\cite{Bern:2021ppb} that in turn was inspired by Ref.~\cite{Caron-Huot:2020cmc}. It is interesting to note that while the accumulation-point model of Ref.~\cite{Bern:2021ppb} was designed to violate low-spin dominance, the associated low-energy EFT still belongs to the same theory island as the more conventional string- and field-theory data. It would be interesting to study the ultimate fate of accumulation-point amplitudes and determine whether or not they should be thought of as physical UV completions. 

Along similar lines, it would be interesting to think about additional models that UV-complete gravitational scattering.  Perhaps the most interesting new classes of theories involves nonperturbative physics. Here, we studied only a first example. It is possible to carry out a controlled non-perturbative analysis in the context of the AdS/CFT correspondence~\cite{Maldacena:1997re, Witten:1998qj, Gubser:1998bc}, generalized as in e.g.~Ref.~\cite{Polchinski:2000uf} or with a simple cutoff in the transverse direction~\cite{Polchinski:2001tt}, to describe a confining boundary theory. 
Indeed, bulk AdS$_5$ supergravity can be used to find the boundary four-dimensional correlation functions of e.g. four stress tensors, which in turn can be interpreted as the four-graviton off-shell Green's functions. Fourier-transforming to boundary momentum space allows on-shell conditions to be imposed and leads to the boundary four-graviton amplitude due to virtual nonperturbative matter.
Naively, the contribution of one bulk-exchange diagram to the amplitude is, up to a polynomial helicity-dependent factor, a function of the corresponding Mandelstam invariant. It would be very interesting to explore the consequences of these 
polynomial factors on the Taylor coefficients of the expansion of the amplitude 
at large confinement scale.\footnote{At high energies the scattering process is 
localized in AdS space and given, up to external-state factors, by tree-level string theory amplitudes~\cite{Polchinski:2001tt}. This suggests---but does not prove---that we may expect certain similarities between the properties of flat-space string-theory Taylor coefficients and those of the boundary graviton amplitudes.}

In conclusion, placing bounds on gravitational scattering and analyzing explicit data in four spacetime dimensions and beyond should give us a fruitful probe of possible extensions of Einstein gravity.  A key question is whether we can  constrain sensible theories to live on islands as small as those
suggested by the explicit theoretical data.

\subsection*{Note added:} 
While completing this manuscript, we were informed by Simon Caron-Huot, David Simmons-Duffin, Julio Parra-Martinez, and Yue Zhou about their study of ``Graviton partial waves and causality in higher dimensions'' \cite{Caron-Huot:2022toAppear}. It would be interesting to compare the explicit data presented here to the bounded regions derived from the dispersive arguments and study possible low-spin dominance explanations of the small theory islands. 

\section*{Acknowledgments:} 
%
We are grateful to Simon Caron-Huot, David Simmons-Duffin, Julio Parra-Martinez, and Yue Zhou for pointing out their upcoming work on higher-dimensional spinning partial waves to us prior to publication~\cite{Caron-Huot:2022toAppear}. We thank Petr Kravchuk and Alexander Zhiboedov for enlightening discussions and collaboration on related topics. We further wish to thank Callum Jones, Mikhail Solon and Fei Teng for fruitful conversations. Z.B, E.H. and D.K. are supported by the U.S.\ Department of Energy (DOE) under Award Number DE-SC0009937. R.R. is supported by the U.S.  Department of Energy (DOE) under award number~DE-SC00019066. We also are grateful to the Mani L. Bhaumik Institute for Theoretical Physics for support. 


\bibliographystyle{JHEP}
\bibliography{refs_eft_data.bib}

\providecommand{\href}[2]{#2}\begingroup\raggedright\begin{thebibliography}{10}

\bibitem{Buchmuller:2001dc}
W.~Buchmuller and D.~Wyler, \emph{{CP violation, neutrino mixing and the baryon
  asymmetry}}, \href{https://doi.org/10.1016/S0370-2693(01)01211-4}{\emph{Phys.
  Lett. B} {\bfseries 521} (2001) 291}
  [\href{https://arxiv.org/abs/hep-ph/0108216}{{\ttfamily hep-ph/0108216}}].

\bibitem{Brivio:2017vri}
I.~Brivio and M.~Trott, \emph{{The Standard Model as an Effective Field
  Theory}}, \href{https://doi.org/10.1016/j.physrep.2018.11.002}{\emph{Phys.
  Rept.} {\bfseries 793} (2019) 1}
  [\href{https://arxiv.org/abs/1706.08945}{{\ttfamily 1706.08945}}].

\bibitem{Dubovsky:2011sj}
S.~Dubovsky, L.~Hui, A.~Nicolis and D.~T. Son, \emph{{Effective field theory
  for hydrodynamics: thermodynamics, and the derivative expansion}},
  \href{https://doi.org/10.1103/PhysRevD.85.085029}{\emph{Phys. Rev. D}
  {\bfseries 85} (2012) 085029}
  [\href{https://arxiv.org/abs/1107.0731}{{\ttfamily 1107.0731}}].

\bibitem{Cheung:2007st}
C.~Cheung, P.~Creminelli, A.~L. Fitzpatrick, J.~Kaplan and L.~Senatore,
  \emph{{The Effective Field Theory of Inflation}},
  \href{https://doi.org/10.1088/1126-6708/2008/03/014}{\emph{JHEP} {\bfseries
  03} (2008) 014} [\href{https://arxiv.org/abs/0709.0293}{{\ttfamily
  0709.0293}}].

\bibitem{Carrasco:2012cv}
J.~J.~M. Carrasco, M.~P. Hertzberg and L.~Senatore, \emph{{The Effective Field
  Theory of Cosmological Large Scale Structures}},
  \href{https://doi.org/10.1007/JHEP09(2012)082}{\emph{JHEP} {\bfseries 09}
  (2012) 082} [\href{https://arxiv.org/abs/1206.2926}{{\ttfamily 1206.2926}}].

\bibitem{Goldberger:2004jt}
W.~D. Goldberger and I.~Z. Rothstein, \emph{{An Effective field theory of
  gravity for extended objects}},
  \href{https://doi.org/10.1103/PhysRevD.73.104029}{\emph{Phys. Rev. D}
  {\bfseries 73} (2006) 104029}
  [\href{https://arxiv.org/abs/hep-th/0409156}{{\ttfamily hep-th/0409156}}].

\bibitem{Adams:2006sv}
A.~Adams, N.~Arkani-Hamed, S.~Dubovsky, A.~Nicolis and R.~Rattazzi,
  \emph{{Causality, analyticity and an IR obstruction to UV completion}},
  \href{https://doi.org/10.1088/1126-6708/2006/10/014}{\emph{JHEP} {\bfseries
  10} (2006) 014} [\href{https://arxiv.org/abs/hep-th/0602178}{{\ttfamily
  hep-th/0602178}}].

\bibitem{Camanho:2014apa}
X.~O. Camanho, J.~D. Edelstein, J.~Maldacena and A.~Zhiboedov, \emph{{Causality
  Constraints on Corrections to the Graviton Three-Point Coupling}},
  \href{https://doi.org/10.1007/JHEP02(2016)020}{\emph{JHEP} {\bfseries 02}
  (2016) 020} [\href{https://arxiv.org/abs/1407.5597}{{\ttfamily 1407.5597}}].

\bibitem{Froissart:1961ux}
M.~Froissart, \emph{{Asymptotic behavior and subtractions in the Mandelstam
  representation}}, \href{https://doi.org/10.1103/PhysRev.123.1053}{\emph{Phys.
  Rev.} {\bfseries 123} (1961) 1053}.

\bibitem{Martin:1962rt}
A.~Martin, \emph{{Unitarity and high-energy behavior of scattering
  amplitudes}}, \href{https://doi.org/10.1103/PhysRev.129.1432}{\emph{Phys.
  Rev.} {\bfseries 129} (1963) 1432}.

\bibitem{Maldacena:2015waa}
J.~Maldacena, S.~H. Shenker and D.~Stanford, \emph{{A bound on chaos}},
  \href{https://doi.org/10.1007/JHEP08(2016)106}{\emph{JHEP} {\bfseries 08}
  (2016) 106} [\href{https://arxiv.org/abs/1503.01409}{{\ttfamily
  1503.01409}}].

\bibitem{Haring:2022cyf}
K.~H\"aring and A.~Zhiboedov, \emph{{Gravitational Regge bounds}},
  \href{https://arxiv.org/abs/2202.08280}{{\ttfamily 2202.08280}}.

\bibitem{Bellazzini:2020cot}
B.~Bellazzini, J.~Elias~Mir\'o, R.~Rattazzi, M.~Riembau and F.~Riva,
  \emph{{Positive moments for scattering amplitudes}},
  \href{https://doi.org/10.1103/PhysRevD.104.036006}{\emph{Phys. Rev. D}
  {\bfseries 104} (2021) 036006}
  [\href{https://arxiv.org/abs/2011.00037}{{\ttfamily 2011.00037}}].

\bibitem{Arkani-Hamed:2020blm}
N.~Arkani-Hamed, T.-C. Huang and Y.-T. Huang, \emph{{The EFT-Hedron}},
  \href{https://doi.org/10.1007/JHEP05(2021)259}{\emph{JHEP} {\bfseries 05}
  (2021) 259} [\href{https://arxiv.org/abs/2012.15849}{{\ttfamily
  2012.15849}}].

\bibitem{Bern:2021ppb}
Z.~Bern, D.~Kosmopoulos and A.~Zhiboedov, \emph{{Gravitational Effective Field
  Theory Islands, Low-Spin Dominance, and the Four-Graviton Amplitude}},
  \href{https://arxiv.org/abs/2103.12728}{{\ttfamily 2103.12728}}.

\bibitem{Caron-Huot:2021enk}
S.~Caron-Huot, D.~Mazac, L.~Rastelli and D.~Simmons-Duffin, \emph{{AdS bulk
  locality from sharp CFT bounds}},
  \href{https://doi.org/10.1007/JHEP11(2021)164}{\emph{JHEP} {\bfseries 11}
  (2021) 164} [\href{https://arxiv.org/abs/2106.10274}{{\ttfamily
  2106.10274}}].

\bibitem{Caron-Huot:2022ugt}
S.~Caron-Huot, Y.-Z. Li, J.~Parra-Martinez and D.~Simmons-Duffin,
  \emph{{Causality constraints on corrections to Einstein gravity}},
  \href{https://arxiv.org/abs/2201.06602}{{\ttfamily 2201.06602}}.

\bibitem{Chiang:2022jep}
L.-Y. Chiang, Y.-t. Huang, W.~Li, L.~Rodina and H.-C. Weng,
  \emph{{(Non)-projective bounds on gravitational EFT}},
  \href{https://arxiv.org/abs/2201.07177}{{\ttfamily 2201.07177}}.

\bibitem{Chiang:2022ltp}
L.-Y. Chiang, Y.-t. Huang, L.~Rodina and H.-C. Weng, \emph{{De-projecting the
  EFThedron}},  \href{https://arxiv.org/abs/2204.07140}{{\ttfamily
  2204.07140}}.

\bibitem{Shadmi:2018xan}
Y.~Shadmi and Y.~Weiss, \emph{{Effective Field Theory Amplitudes the On-Shell
  Way: Scalar and Vector Couplings to Gluons}},
  \href{https://doi.org/10.1007/JHEP02(2019)165}{\emph{JHEP} {\bfseries 02}
  (2019) 165} [\href{https://arxiv.org/abs/1809.09644}{{\ttfamily
  1809.09644}}].

\bibitem{Ferrara:1972kab}
S.~Ferrara, A.~F. Grillo, G.~Parisi and R.~Gatto, \emph{{Covariant expansion of
  the conformal four-point function}},
  \href{https://doi.org/10.1016/0550-3213(73)90467-7}{\emph{Nucl. Phys. B}
  {\bfseries 49} (1972) 77}.

\bibitem{Ferrara:1974ny}
S.~Ferrara, R.~Gatto and A.~F. Grillo, \emph{{Properties of Partial Wave
  Amplitudes in Conformal Invariant Field Theories}},
  \href{https://doi.org/10.1007/BF02769009}{\emph{Nuovo Cim. A} {\bfseries 26}
  (1975) 226}.

\bibitem{Dolan:2000ut}
F.~A. Dolan and H.~Osborn, \emph{{Conformal four point functions and the
  operator product expansion}},
  \href{https://doi.org/10.1016/S0550-3213(01)00013-X}{\emph{Nucl. Phys. B}
  {\bfseries 599} (2001) 459}
  [\href{https://arxiv.org/abs/hep-th/0011040}{{\ttfamily hep-th/0011040}}].

\bibitem{Dolan:2003hv}
F.~A. Dolan and H.~Osborn, \emph{{Conformal partial waves and the operator
  product expansion}},
  \href{https://doi.org/10.1016/j.nuclphysb.2003.11.016}{\emph{Nucl. Phys. B}
  {\bfseries 678} (2004) 491}
  [\href{https://arxiv.org/abs/hep-th/0309180}{{\ttfamily hep-th/0309180}}].

\bibitem{Itzykson:1980rh}
C.~Itzykson and J.~B. Zuber, \emph{{Quantum Field Theory}}, International
  Series In Pure and Applied Physics. McGraw-Hill, New York, 1980.

\bibitem{Hebbar:2020ukp}
A.~Hebbar, D.~Karateev and J.~Penedones, \emph{{Spinning S-matrix Bootstrap in
  4d}},  \href{https://arxiv.org/abs/2011.11708}{{\ttfamily 2011.11708}}.

\bibitem{Pham:1985cr}
T.~N. Pham and T.~N. Truong, \emph{{Evaluation of the Derivative Quartic Terms
  of the Meson Chiral Lagrangian From Forward Dispersion Relation}},
  \href{https://doi.org/10.1103/PhysRevD.31.3027}{\emph{Phys. Rev. D}
  {\bfseries 31} (1985) 3027}.

\bibitem{Ananthanarayan:1994hf}
B.~Ananthanarayan, D.~Toublan and G.~Wanders, \emph{{Consistency of the chiral
  pion pion scattering amplitudes with axiomatic constraints}},
  \href{https://doi.org/10.1103/PhysRevD.51.1093}{\emph{Phys. Rev. D}
  {\bfseries 51} (1995) 1093}
  [\href{https://arxiv.org/abs/hep-ph/9410302}{{\ttfamily hep-ph/9410302}}].

\bibitem{Pennington:1994kc}
M.~R. Pennington and J.~Portoles, \emph{{The Chiral Lagrangian parameters, l1,
  l2, are determined by the rho resonance}},
  \href{https://doi.org/10.1016/0370-2693(94)01551-M}{\emph{Phys. Lett. B}
  {\bfseries 344} (1995) 399}
  [\href{https://arxiv.org/abs/hep-ph/9409426}{{\ttfamily hep-ph/9409426}}].

\bibitem{Nicolis:2009qm}
A.~Nicolis, R.~Rattazzi and E.~Trincherini, \emph{{Energy's and amplitudes'
  positivity}}, \href{https://doi.org/10.1007/JHEP05(2010)095}{\emph{JHEP}
  {\bfseries 05} (2010) 095} [\href{https://arxiv.org/abs/0912.4258}{{\ttfamily
  0912.4258}}].

\bibitem{Bellazzini:2015cra}
B.~Bellazzini, C.~Cheung and G.~N. Remmen, \emph{{Quantum Gravity Constraints
  from Unitarity and Analyticity}},
  \href{https://doi.org/10.1103/PhysRevD.93.064076}{\emph{Phys. Rev. D}
  {\bfseries 93} (2016) 064076}
  [\href{https://arxiv.org/abs/1509.00851}{{\ttfamily 1509.00851}}].

\bibitem{deRham:2017avq}
C.~de~Rham, S.~Melville, A.~J. Tolley and S.-Y. Zhou, \emph{{Positivity bounds
  for scalar field theories}},
  \href{https://doi.org/10.1103/PhysRevD.96.081702}{\emph{Phys. Rev. D}
  {\bfseries 96} (2017) 081702}
  [\href{https://arxiv.org/abs/1702.06134}{{\ttfamily 1702.06134}}].

\bibitem{deRham:2017zjm}
C.~de~Rham, S.~Melville, A.~J. Tolley and S.-Y. Zhou, \emph{{UV complete me:
  Positivity Bounds for Particles with Spin}},
  \href{https://doi.org/10.1007/JHEP03(2018)011}{\emph{JHEP} {\bfseries 03}
  (2018) 011} [\href{https://arxiv.org/abs/1706.02712}{{\ttfamily
  1706.02712}}].

\bibitem{Sinha:2020win}
A.~Sinha and A.~Zahed, \emph{{Crossing Symmetric Dispersion Relations in
  Quantum Field Theories}},
  \href{https://doi.org/10.1103/PhysRevLett.126.181601}{\emph{Phys. Rev. Lett.}
  {\bfseries 126} (2021) 181601}
  [\href{https://arxiv.org/abs/2012.04877}{{\ttfamily 2012.04877}}].

\bibitem{Chowdhury:2021ynh}
S.~D. Chowdhury, K.~Ghosh, P.~Haldar, P.~Raman and A.~Sinha, \emph{{Crossing
  Symmetric Spinning S-matrix Bootstrap: EFT bounds}},
  \href{https://arxiv.org/abs/2112.11755}{{\ttfamily 2112.11755}}.

\bibitem{Bellazzini:2021oaj}
B.~Bellazzini, M.~Riembau and F.~Riva, \emph{{The IR-Side of Positivity
  Bounds}},  \href{https://arxiv.org/abs/2112.12561}{{\ttfamily 2112.12561}}.

\bibitem{Chiang:2021ziz}
L.-Y. Chiang, Y.-t. Huang, W.~Li, L.~Rodina and H.-C. Weng, \emph{{Into the
  EFThedron and UV constraints from IR consistency}},
  \href{https://arxiv.org/abs/2105.02862}{{\ttfamily 2105.02862}}.

\bibitem{Cheung:2014ega}
C.~Cheung and G.~N. Remmen, \emph{{Infrared Consistency and the Weak Gravity
  Conjecture}}, \href{https://doi.org/10.1007/JHEP12(2014)087}{\emph{JHEP}
  {\bfseries 12} (2014) 087} [\href{https://arxiv.org/abs/1407.7865}{{\ttfamily
  1407.7865}}].

\bibitem{Cheung:2014vva}
C.~Cheung and G.~N. Remmen, \emph{{Naturalness and the Weak Gravity
  Conjecture}},
  \href{https://doi.org/10.1103/PhysRevLett.113.051601}{\emph{Phys. Rev. Lett.}
  {\bfseries 113} (2014) 051601}
  [\href{https://arxiv.org/abs/1402.2287}{{\ttfamily 1402.2287}}].

\bibitem{Arkani-Hamed:2021ajd}
N.~Arkani-Hamed, Y.-t. Huang, J.-Y. Liu and G.~N. Remmen, \emph{{Causality,
  Unitarity, and the Weak Gravity Conjecture}},
  \href{https://arxiv.org/abs/2109.13937}{{\ttfamily 2109.13937}}.

\bibitem{Caron-Huot:2020adz}
S.~Caron-Huot, D.~Mazac, L.~Rastelli and D.~Simmons-Duffin, \emph{{Dispersive
  CFT Sum Rules}}, \href{https://doi.org/10.1007/JHEP05(2021)243}{\emph{JHEP}
  {\bfseries 05} (2021) 243}
  [\href{https://arxiv.org/abs/2008.04931}{{\ttfamily 2008.04931}}].

\bibitem{Distler:2006if}
J.~Distler, B.~Grinstein, R.~A. Porto and I.~Z. Rothstein, \emph{{Falsifying
  Models of New Physics via WW Scattering}},
  \href{https://doi.org/10.1103/PhysRevLett.98.041601}{\emph{Phys. Rev. Lett.}
  {\bfseries 98} (2007) 041601}
  [\href{https://arxiv.org/abs/hep-ph/0604255}{{\ttfamily hep-ph/0604255}}].

\bibitem{Manohar:2008tc}
A.~V. Manohar and V.~Mateu, \emph{{Dispersion Relation Bounds for pi pi
  Scattering}}, \href{https://doi.org/10.1103/PhysRevD.77.094019}{\emph{Phys.
  Rev. D} {\bfseries 77} (2008) 094019}
  [\href{https://arxiv.org/abs/0801.3222}{{\ttfamily 0801.3222}}].

\bibitem{Remmen:2019cyz}
G.~N. Remmen and N.~L. Rodd, \emph{{Consistency of the Standard Model Effective
  Field Theory}}, \href{https://doi.org/10.1007/JHEP12(2019)032}{\emph{JHEP}
  {\bfseries 12} (2019) 032}
  [\href{https://arxiv.org/abs/1908.09845}{{\ttfamily 1908.09845}}].

\bibitem{Remmen:2020uze}
G.~N. Remmen and N.~L. Rodd, \emph{{Signs, spin, SMEFT: Sum rules at dimension
  six}}, \href{https://doi.org/10.1103/PhysRevD.105.036006}{\emph{Phys. Rev. D}
  {\bfseries 105} (2022) 036006}
  [\href{https://arxiv.org/abs/2010.04723}{{\ttfamily 2010.04723}}].

\bibitem{Caron-Huot:2020cmc}
S.~Caron-Huot and V.~Van~Duong, \emph{{Extremal Effective Field Theories}},
  \href{https://doi.org/10.1007/JHEP05(2021)280}{\emph{JHEP} {\bfseries 05}
  (2021) 280} [\href{https://arxiv.org/abs/2011.02957}{{\ttfamily
  2011.02957}}].

\bibitem{Caron-Huot:2021rmr}
S.~Caron-Huot, D.~Mazac, L.~Rastelli and D.~Simmons-Duffin, \emph{{Sharp
  Boundaries for the Swampland}},
  \href{https://doi.org/10.1007/jhep07(2021)110}{\emph{JHEP} {\bfseries 07}
  (2021) 110} [\href{https://arxiv.org/abs/2102.08951}{{\ttfamily
  2102.08951}}].

\bibitem{Cheung:2016yqr}
C.~Cheung and G.~N. Remmen, \emph{{Positive Signs in Massive Gravity}},
  \href{https://doi.org/10.1007/JHEP04(2016)002}{\emph{JHEP} {\bfseries 04}
  (2016) 002} [\href{https://arxiv.org/abs/1601.04068}{{\ttfamily
  1601.04068}}].

\bibitem{Zhang:2020jyn}
C.~Zhang and S.-Y. Zhou, \emph{{Convex Geometry Perspective on the (Standard
  Model) Effective Field Theory Space}},
  \href{https://doi.org/10.1103/PhysRevLett.125.201601}{\emph{Phys. Rev. Lett.}
  {\bfseries 125} (2020) 201601}
  [\href{https://arxiv.org/abs/2005.03047}{{\ttfamily 2005.03047}}].

\bibitem{Li:2021lpe}
X.~Li, H.~Xu, C.~Yang, C.~Zhang and S.-Y. Zhou, \emph{{Positivity in Multifield
  Effective Field Theories}},
  \href{https://doi.org/10.1103/PhysRevLett.127.121601}{\emph{Phys. Rev. Lett.}
  {\bfseries 127} (2021) 121601}
  [\href{https://arxiv.org/abs/2101.01191}{{\ttfamily 2101.01191}}].

\bibitem{Caron-Huot:2022toAppear}
S.~Caron-Huot, D.~Simmons-Duffin, J.~Parra-Martinez and Y.~Zhou,
  \emph{{Graviton partial waves and causality in higher dimensions}}, {\emph{to
  appear} (2022) } [\href{https://arxiv.org/abs/2205.xxxxx}{{\ttfamily
  2205.xxxxx}}].

\bibitem{Guerrieri:2021ivu}
A.~Guerrieri, J.~Penedones and P.~Vieira, \emph{{Where is String Theory?}},
  \href{https://arxiv.org/abs/2102.02847}{{\ttfamily 2102.02847}}.

\bibitem{Boels:2009bv}
R.~Boels, \emph{{Covariant representation theory of the Poincare algebra and
  some of its extensions}},
  \href{https://doi.org/10.1007/JHEP01(2010)010}{\emph{JHEP} {\bfseries 01}
  (2010) 010} [\href{https://arxiv.org/abs/0908.0738}{{\ttfamily 0908.0738}}].

\bibitem{Chowdhury:2019kaq}
S.~D. Chowdhury, A.~Gadde, T.~Gopalka, I.~Halder, L.~Janagal and S.~Minwalla,
  \emph{{Classifying and constraining local four photon and four graviton
  S-matrices}}, \href{https://doi.org/10.1007/JHEP02(2020)114}{\emph{JHEP}
  {\bfseries 02} (2020) 114}
  [\href{https://arxiv.org/abs/1910.14392}{{\ttfamily 1910.14392}}].

\bibitem{Grisaru:1979wc}
M.~T. Grisaru, W.~Siegel and M.~Rocek, \emph{{Improved Methods for
  Supergraphs}},
  \href{https://doi.org/10.1016/0550-3213(79)90344-4}{\emph{Nucl. Phys. B}
  {\bfseries 159} (1979) 429}.

\bibitem{Seiberg:1994bz}
N.~Seiberg, \emph{{Exact results on the space of vacua of four-dimensional SUSY
  gauge theories}}, \href{https://doi.org/10.1103/PhysRevD.49.6857}{\emph{Phys.
  Rev. D} {\bfseries 49} (1994) 6857}
  [\href{https://arxiv.org/abs/hep-th/9402044}{{\ttfamily hep-th/9402044}}].

\bibitem{Intriligator:1994jr}
K.~A. Intriligator, R.~G. Leigh and N.~Seiberg, \emph{{Exact superpotentials in
  four-dimensions}},
  \href{https://doi.org/10.1103/PhysRevD.50.1092}{\emph{Phys. Rev. D}
  {\bfseries 50} (1994) 1092}
  [\href{https://arxiv.org/abs/hep-th/9403198}{{\ttfamily hep-th/9403198}}].

\bibitem{Affleck:1983vc}
I.~Affleck, M.~Dine and N.~Seiberg, \emph{{Dynamical Supersymmetry Breaking in
  Chiral Theories}},
  \href{https://doi.org/10.1016/0370-2693(84)90227-2}{\emph{Phys. Lett. B}
  {\bfseries 137} (1984) 187}.

\bibitem{Dijkgraaf:2002dh}
R.~Dijkgraaf and C.~Vafa, \emph{{A Perturbative window into nonperturbative
  physics}},  \href{https://arxiv.org/abs/hep-th/0208048}{{\ttfamily
  hep-th/0208048}}.

\bibitem{Dijkgraaf:2002fc}
R.~Dijkgraaf and C.~Vafa, \emph{{Matrix models, topological strings, and
  supersymmetric gauge theories}},
  \href{https://doi.org/10.1016/S0550-3213(02)00766-6}{\emph{Nucl. Phys. B}
  {\bfseries 644} (2002) 3}
  [\href{https://arxiv.org/abs/hep-th/0206255}{{\ttfamily hep-th/0206255}}].

\bibitem{Bena:2002kw}
I.~Bena and R.~Roiban, \emph{{Exact superpotentials in N = 1 theories with
  flavor and their matrix model formulation}},
  \href{https://doi.org/10.1016/S0370-2693(03)00034-0}{\emph{Phys. Lett. B}
  {\bfseries 555} (2003) 117}
  [\href{https://arxiv.org/abs/hep-th/0211075}{{\ttfamily hep-th/0211075}}].

\bibitem{Dijkgraaf:2002xd}
R.~Dijkgraaf, M.~T. Grisaru, C.~S. Lam, C.~Vafa and D.~Zanon,
  \emph{{Perturbative computation of glueball superpotentials}},
  \href{https://doi.org/10.1016/j.physletb.2003.08.060}{\emph{Phys. Lett. B}
  {\bfseries 573} (2003) 138}
  [\href{https://arxiv.org/abs/hep-th/0211017}{{\ttfamily hep-th/0211017}}].

\bibitem{Dijkgraaf:2003sk}
R.~Dijkgraaf, M.~T. Grisaru, H.~Ooguri, C.~Vafa and D.~Zanon, \emph{{Planar
  gravitational corrections for supersymmetric gauge theories}},
  \href{https://doi.org/10.1088/1126-6708/2004/04/028}{\emph{JHEP} {\bfseries
  04} (2004) 028} [\href{https://arxiv.org/abs/hep-th/0310061}{{\ttfamily
  hep-th/0310061}}].

\bibitem{Cachazo:2002ry}
F.~Cachazo, M.~R. Douglas, N.~Seiberg and E.~Witten, \emph{{Chiral rings and
  anomalies in supersymmetric gauge theory}},
  \href{https://doi.org/10.1088/1126-6708/2002/12/071}{\emph{JHEP} {\bfseries
  12} (2002) 071} [\href{https://arxiv.org/abs/hep-th/0211170}{{\ttfamily
  hep-th/0211170}}].

\bibitem{Cachazo:2002zk}
F.~Cachazo, N.~Seiberg and E.~Witten, \emph{{Phases of N=1 supersymmetric gauge
  theories and matrices}},
  \href{https://doi.org/10.1088/1126-6708/2003/02/042}{\emph{JHEP} {\bfseries
  02} (2003) 042} [\href{https://arxiv.org/abs/hep-th/0301006}{{\ttfamily
  hep-th/0301006}}].

\bibitem{Cachazo:2003yc}
F.~Cachazo, N.~Seiberg and E.~Witten, \emph{{Chiral rings and phases of
  supersymmetric gauge theories}},
  \href{https://doi.org/10.1088/1126-6708/2003/04/018}{\emph{JHEP} {\bfseries
  04} (2003) 018} [\href{https://arxiv.org/abs/hep-th/0303207}{{\ttfamily
  hep-th/0303207}}].

\bibitem{Veneziano:1982ah}
G.~Veneziano and S.~Yankielowicz, \emph{{An Effective Lagrangian for the Pure
  N=1 Supersymmetric Yang-Mills Theory}},
  \href{https://doi.org/10.1016/0370-2693(82)90828-0}{\emph{Phys. Lett. B}
  {\bfseries 113} (1982) 231}.

\bibitem{Kawai:1985xq}
H.~Kawai, D.~C. Lewellen and S.~H.~H. Tye, \emph{{A Relation Between Tree
  Amplitudes of Closed and Open Strings}},
  \href{https://doi.org/10.1016/0550-3213(86)90362-7}{\emph{Nucl. Phys. B}
  {\bfseries 269} (1986) 1}.

\bibitem{Green:1982sw}
M.~B. Green, J.~H. Schwarz and L.~Brink, \emph{{N=4 Yang-Mills and N=8
  Supergravity as Limits of String Theories}},
  \href{https://doi.org/10.1016/0550-3213(82)90336-4}{\emph{Nucl. Phys. B}
  {\bfseries 198} (1982) 474}.

\bibitem{Bern:1994zx}
Z.~Bern, L.~J. Dixon, D.~C. Dunbar and D.~A. Kosower, \emph{{One loop n point
  gauge theory amplitudes, unitarity and collinear limits}},
  \href{https://doi.org/10.1016/0550-3213(94)90179-1}{\emph{Nucl. Phys. B}
  {\bfseries 425} (1994) 217}
  [\href{https://arxiv.org/abs/hep-ph/9403226}{{\ttfamily hep-ph/9403226}}].

\bibitem{Bern:1994cg}
Z.~Bern, L.~J. Dixon, D.~C. Dunbar and D.~A. Kosower, \emph{{Fusing gauge
  theory tree amplitudes into loop amplitudes}},
  \href{https://doi.org/10.1016/0550-3213(94)00488-Z}{\emph{Nucl. Phys. B}
  {\bfseries 435} (1995) 59}
  [\href{https://arxiv.org/abs/hep-ph/9409265}{{\ttfamily hep-ph/9409265}}].

\bibitem{Britto:2004nc}
R.~Britto, F.~Cachazo and B.~Feng, \emph{{Generalized unitarity and one-loop
  amplitudes in N=4 super-Yang-Mills}},
  \href{https://doi.org/10.1016/j.nuclphysb.2005.07.014}{\emph{Nucl. Phys. B}
  {\bfseries 725} (2005) 275}
  [\href{https://arxiv.org/abs/hep-th/0412103}{{\ttfamily hep-th/0412103}}].

\bibitem{Bern:2020buy}
Z.~Bern, A.~Luna, R.~Roiban, C.-H. Shen and M.~Zeng, \emph{{Spinning Black Hole
  Binary Dynamics, Scattering Amplitudes and Effective Field Theory}},
  \href{https://arxiv.org/abs/2005.03071}{{\ttfamily 2005.03071}}.

\bibitem{Smirnov:2012gma}
V.~A. Smirnov, \emph{{Analytic tools for Feynman integrals}}, vol.~250.
  Springer, 2012,
  \href{https://doi.org/10.1007/978-3-642-34886-0}{10.1007/978-3-642-34886-0}.

\bibitem{Tarasov:1996bz}
O.~V. Tarasov, \emph{{A New approach to the momentum expansion of multiloop
  Feynman diagrams}},
  \href{https://doi.org/10.1016/S0550-3213(96)00466-X}{\emph{Nucl. Phys. B}
  {\bfseries 480} (1996) 397}
  [\href{https://arxiv.org/abs/hep-ph/9606238}{{\ttfamily hep-ph/9606238}}].

\bibitem{Tarasov:1997kx}
O.~V. Tarasov, \emph{{Generalized recurrence relations for two loop propagator
  integrals with arbitrary masses}},
  \href{https://doi.org/10.1016/S0550-3213(97)00376-3}{\emph{Nucl. Phys. B}
  {\bfseries 502} (1997) 455}
  [\href{https://arxiv.org/abs/hep-ph/9703319}{{\ttfamily hep-ph/9703319}}].

\bibitem{Tarasov:1996br}
O.~V. Tarasov, \emph{{Connection between Feynman integrals having different
  values of the space-time dimension}},
  \href{https://doi.org/10.1103/PhysRevD.54.6479}{\emph{Phys. Rev. D}
  {\bfseries 54} (1996) 6479}
  [\href{https://arxiv.org/abs/hep-th/9606018}{{\ttfamily hep-th/9606018}}].

\bibitem{Lee:2009dh}
R.~N. Lee, \emph{{Space-time dimensionality D as complex variable: Calculating
  loop integrals using dimensional recurrence relation and analytical
  properties with respect to D}},
  \href{https://doi.org/10.1016/j.nuclphysb.2009.12.025}{\emph{Nucl. Phys. B}
  {\bfseries 830} (2010) 474}
  [\href{https://arxiv.org/abs/0911.0252}{{\ttfamily 0911.0252}}].

\bibitem{Lee:2010wea}
R.~N. Lee, \emph{{Calculating multiloop integrals using dimensional recurrence
  relation and $D$-analyticity}},
  \href{https://doi.org/10.1016/j.nuclphysbps.2010.08.032}{\emph{Nucl. Phys. B
  Proc. Suppl.} {\bfseries 205-206} (2010) 135}
  [\href{https://arxiv.org/abs/1007.2256}{{\ttfamily 1007.2256}}].

\bibitem{Anastasiou:2005cb}
C.~Anastasiou and A.~Daleo, \emph{{Numerical evaluation of loop integrals}},
  \href{https://doi.org/10.1088/1126-6708/2006/10/031}{\emph{JHEP} {\bfseries
  10} (2006) 031} [\href{https://arxiv.org/abs/hep-ph/0511176}{{\ttfamily
  hep-ph/0511176}}].

\bibitem{Tkachov:1981wb}
F.~V. Tkachov, \emph{{A Theorem on Analytical Calculability of Four Loop
  Renormalization Group Functions}},
  \href{https://doi.org/10.1016/0370-2693(81)90288-4}{\emph{Phys. Lett. B}
  {\bfseries 100} (1981) 65}.

\bibitem{Chetyrkin:1981qh}
K.~Chetyrkin and F.~Tkachov, \emph{{Integration by Parts: The Algorithm to
  Calculate beta Functions in 4 Loops}},
  \href{https://doi.org/10.1016/0550-3213(81)90199-1}{\emph{Nucl. Phys. B}
  {\bfseries 192} (1981) 159}.

\bibitem{Laporta:1996mq}
S.~Laporta and E.~Remiddi, \emph{{The Analytical value of the electron (g-2) at
  order alpha**3 in QED}},
  \href{https://doi.org/10.1016/0370-2693(96)00439-X}{\emph{Phys. Lett. B}
  {\bfseries 379} (1996) 283}
  [\href{https://arxiv.org/abs/hep-ph/9602417}{{\ttfamily hep-ph/9602417}}].

\bibitem{Laporta:2001dd}
S.~Laporta, \emph{{High precision calculation of multiloop Feynman integrals by
  difference equations}},
  \href{https://doi.org/10.1016/S0217-751X(00)00215-7}{\emph{Int. J. Mod. Phys.
  A} {\bfseries 15} (2000) 5087}
  [\href{https://arxiv.org/abs/hep-ph/0102033}{{\ttfamily hep-ph/0102033}}].

\bibitem{Smirnov:2008iw}
A.~V. Smirnov, \emph{{Algorithm FIRE -- Feynman Integral REduction}},
  \href{https://doi.org/10.1088/1126-6708/2008/10/107}{\emph{JHEP} {\bfseries
  10} (2008) 107} [\href{https://arxiv.org/abs/0807.3243}{{\ttfamily
  0807.3243}}].

\bibitem{Smirnov:2019qkx}
A.~V. Smirnov and F.~S. Chuharev, \emph{{FIRE6: Feynman Integral REduction with
  Modular Arithmetic}},
  \href{https://doi.org/10.1016/j.cpc.2019.106877}{\emph{Comput. Phys. Commun.}
  {\bfseries 247} (2020) 106877}
  [\href{https://arxiv.org/abs/1901.07808}{{\ttfamily 1901.07808}}].

\bibitem{Bern:1995db}
Z.~Bern and A.~G. Morgan, \emph{{Massive loop amplitudes from unitarity}},
  \href{https://doi.org/10.1016/0550-3213(96)00078-8}{\emph{Nucl. Phys. B}
  {\bfseries 467} (1996) 479}
  [\href{https://arxiv.org/abs/hep-ph/9511336}{{\ttfamily hep-ph/9511336}}].

\bibitem{Ellis:2007qk}
R.~K. Ellis and G.~Zanderighi, \emph{{Scalar one-loop integrals for QCD}},
  \href{https://doi.org/10.1088/1126-6708/2008/02/002}{\emph{JHEP} {\bfseries
  02} (2008) 002} [\href{https://arxiv.org/abs/0712.1851}{{\ttfamily
  0712.1851}}].

\bibitem{Beneke:1997zp}
M.~Beneke and V.~A. Smirnov, \emph{{Asymptotic expansion of Feynman integrals
  near threshold}},
  \href{https://doi.org/10.1016/S0550-3213(98)00138-2}{\emph{Nucl. Phys. B}
  {\bfseries 522} (1998) 321}
  [\href{https://arxiv.org/abs/hep-ph/9711391}{{\ttfamily hep-ph/9711391}}].

\bibitem{Chandorkar:2021viw}
D.~Chandorkar, S.~D. Chowdhury, S.~Kundu and S.~Minwalla, \emph{{Bounds on
  Regge growth of flat space scattering from bounds on chaos}},
  \href{https://doi.org/10.1007/JHEP05(2021)143}{\emph{JHEP} {\bfseries 05}
  (2021) 143} [\href{https://arxiv.org/abs/2102.03122}{{\ttfamily
  2102.03122}}].

\bibitem{Edison:2021ebi}
A.~Edison, M.~Guillen, H.~Johansson, O.~Schlotterer and F.~Teng,
  \emph{{One-loop matrix elements of effective superstring interactions:
  \ensuremath{\alpha}'-expanding loop integrands}},
  \href{https://doi.org/10.1007/JHEP12(2021)007}{\emph{JHEP} {\bfseries 12}
  (2021) 007} [\href{https://arxiv.org/abs/2107.08009}{{\ttfamily
  2107.08009}}].

\bibitem{Figueroa:2022onw}
F.~Figueroa and P.~Tourkine, \emph{{On the unitarity and low energy expansion
  of the Coon amplitude}},  \href{https://arxiv.org/abs/2201.12331}{{\ttfamily
  2201.12331}}.

\bibitem{Huang:2022mdb}
Y.-t. Huang and G.~N. Remmen, \emph{{UV-Complete Gravity Amplitudes and the
  Triple Product}},  \href{https://arxiv.org/abs/2203.00696}{{\ttfamily
  2203.00696}}.

\bibitem{Maldacena:1997re}
J.~M. Maldacena, \emph{{The Large N limit of superconformal field theories and
  supergravity}}, \href{https://doi.org/10.1023/A:1026654312961}{\emph{Adv.
  Theor. Math. Phys.} {\bfseries 2} (1998) 231}
  [\href{https://arxiv.org/abs/hep-th/9711200}{{\ttfamily hep-th/9711200}}].

\bibitem{Witten:1998qj}
E.~Witten, \emph{{Anti-de Sitter space and holography}},
  \href{https://doi.org/10.4310/ATMP.1998.v2.n2.a2}{\emph{Adv. Theor. Math.
  Phys.} {\bfseries 2} (1998) 253}
  [\href{https://arxiv.org/abs/hep-th/9802150}{{\ttfamily hep-th/9802150}}].

\bibitem{Gubser:1998bc}
S.~S. Gubser, I.~R. Klebanov and A.~M. Polyakov, \emph{{Gauge theory
  correlators from noncritical string theory}},
  \href{https://doi.org/10.1016/S0370-2693(98)00377-3}{\emph{Phys. Lett. B}
  {\bfseries 428} (1998) 105}
  [\href{https://arxiv.org/abs/hep-th/9802109}{{\ttfamily hep-th/9802109}}].

\bibitem{Polchinski:2000uf}
J.~Polchinski and M.~J. Strassler, \emph{{The String dual of a confining
  four-dimensional gauge theory}},
  \href{https://arxiv.org/abs/hep-th/0003136}{{\ttfamily hep-th/0003136}}.

\bibitem{Polchinski:2001tt}
J.~Polchinski and M.~J. Strassler, \emph{{Hard scattering and gauge/string
  duality}}, \href{https://doi.org/10.1103/PhysRevLett.88.031601}{\emph{Phys.
  Rev. Lett.} {\bfseries 88} (2002) 031601}
  [\href{https://arxiv.org/abs/hep-th/0109174}{{\ttfamily hep-th/0109174}}].

\end{thebibliography}\endgroup

\end{document}